\documentclass[a4paper,twocolumn]{article}
\usepackage[utf8]{inputenc}
\usepackage[T1]{fontenc}
\usepackage{amsmath, bm, commath, mathtools,amssymb}
\usepackage{hyperref}
\usepackage{cleveref}
\usepackage{siunitx,booktabs,pbox}
\usepackage[table]{xcolor}
\usepackage{authblk}

\usepackage{float}
\usepackage{subfigure}

\usepackage{natbib}
\setcitestyle{numbers}

\hypersetup{colorlinks = true}
	
\begin{document}
	\title{Turbulence and acoustic waves in compressible flows}
	\author{J. Cerretani}
	\affil{%
    Departamento de F\'isica, Facultad de Ciencias Exactas y Naturales, Universidad de Buenos Aires, Ciudad universitaria, 1428 Buenos Aires, Argentina.
    }%
	\author{P. Dmitruk\footnote{pdmitruk@df.uba.ar}}
	\affil{%
    Departamento de F\'isica, Facultad de Ciencias Exactas y Naturales, Universidad de Buenos Aires and IFIBA,
    CONICET, Ciudad universitaria, 1428 Buenos Aires, Argentina.
    }%

\twocolumn[
\begin{@twocolumnfalse}
	\maketitle
	\begin{abstract}
		In this work, direct numerical simulations of the compressible fluid equations 
        in turbulent regimes are performed. The behavior of the flow is either 
        dominated by purely turbulent phenomena or by the generation of sound waves
        in it. 
        Previous studies suggest that three different types of turbulence may happen 
        at the low Mach number limit in polytropic flows: Nearly incompressible (NI),
        modally equipartitioned compressible (MEC) and compressible wave (CW). 
        The distinction 
        between these types of turbulence is investigated here applying different 
        kinds of forcing. Scaling of density fluctuations with Mach number, 
        comparison of 
        the ratio of transverse velocity fluctuations to longitudinal fluctuations 
        and spectral decomposition of fluctuations are used to distinguish
         the nature of these solutions. 
        From the study of the spatio-temporal spectra and correlation 
        times we quantify the contribution of the waves to the 
        total energy of the system.
        Also, in the dynamics of a compressible flow, three associated 
        correlation times are considered: the non-linear time of local interaction 
        between scales, the sweeping time or non-local time of large scales 
        on small scales, and the time associated with acoustic waves (sound). 
        We observed that different correlation times dominate depending on
        the wave number ($k$), the Mach number and the type of forcing. 
		\vspace{1.5cm}
	\end{abstract}
	\end{@twocolumnfalse}
]

\section{\label{intro}Introduction}
\footnotetext{*pdmitruk@df.uba.ar}

The theory 
of compressible turbulent flows 
has not been developed as much as incompressible turbulence flow models. 
For this reason, the theoretical work focused almost exclusively on the study of 
incompressible turbulence \cite{Batchelor, 2Drur, MIT, Herring}.
One of the aspects of interest that motivates us to study turbulence
in compressible flows
are density fluctuations, naturally absent in the incompressible case.

Compressible flows can be classified in several ways, 
the most common uses the Mach number (M) as a parameter. 
The problems of incompressible flows are governed only by two unknowns, these are the pressure 
and velocity that are obtained from the equations that describe the conservation of the mass and 
the linear momentum, where the density of the fluid is constant.
However, in the compressible flows, the density and temperature of the gas also become variable 
and additional equations are needed to solve the system. These equations are: a 
state equation for the gas and an energy balance equation. The simple gas law is the appropriate 
state equation for most dynamic problems.

An area of interest in turbulent compressible flows is the generation 
and propagation of sound, with the theoretical models 
of Ligthill and Proudman (1952) \cite{Lighthill,Proudman}. Numerical study 
on this problem has been carried out in the work of 
Rubinstein and Zhou (2000) \cite{Rubinstein}. 
Acoustics works \cite{Lighthill} indicate that the compressible and incompressible movements of a turbulent flow are interrelated and it is expected, 
due to the dimensional arguments that suggest that a compressible flow 
behaves dynamically as a compressible flow for low Mach values \cite{Landau} , 
that this relation continues to be valid for turbulent flows of 
subsonic Mach numbers.

The equations of motion for low mach number flows have been recently described \cite{Matthaeus-NI-MHD, Zank-NI-I}, with an equation of state, generalizing 
the equations of Klainerman and Majda \cite{Klain-Majda-1, Klain-Majda-2}. In these formalisms the flow becomes almost incompressible, since the solutions 
expressed as powers of the Mach number are solutions of the incompressible equations.

The works of Klainerman, Majda \cite{Klain-Majda-1, Majda} and Kreiss \cite{Heinz-Otto}, allow us to understand how the solutions of compressible flow 
equations approach, for small mach values, the solutions of incompressible flows.
In all cases the pressure variations must scale as the Mach number squared. 
In this paper, we will refer to the models of Klainerman and Majda \cite{Klain-Majda-1, Klain-Majda-2} 
as theory of nearly incompressible turbulence (NI). For weakly 
compressible flows (Mach numbers less than unity), the theoretical model 
of nearly-incompressible flows (NI) \cite{Zank-NI-I,Ghosh-Relaxation} 
has not been studied numerically for a three-dimensional system. 
In particular, the analysis of the scaling of the fluctuations of 
density with the Mach number is of interest for 
the characterization of the flow. 

Turbulence studies \cite{Pouquet} for Mach numbers less than about $0.3$ infer that one consequence of the NI turbulence theory is that many of the incompressible turbulence characteristics survive in low Mach number flows. Furthermore, simulations of compressible MHD flows showed similarities, especially for low Mach numbers \cite{Shebalin,Dahlburg-1,Dahlburg-2,Dahlburg-3,Scholer,Ghosh-Relaxation}, with the physics of compressible and incompressible flows.

When the nearly incompressible asymptotic expansion NI breaks down, the flow enters in a new regime \cite{Majda, Pouquet}. 
The theoretical work on this new regime of density and velocity fluctuations is described by the Kraichnan \cite{MEC} model. 
The compressible hydrodynamic turbulence model of Kraichnan assumes the existence of strong acoustic waves that are 
energetically equipartitioned with vorticity movements in each scale.
We will refer to this description as modally equipartitioned 
compressible (MEC) turbulence. Kraichnan arrives at a 
different scaling of density and velocity fluctuations with  
the Mach number. 
As seen, in the turbulence model NI, the density fluctuations scale quadratically with the Mach number. On the other hand, 
in the MEC turbulence model, the flows, with larger fluctuations of 
the initial density, become strongly compressible 
and the fluctuations of the density scale linearly with the Mach number. \cite{Ghosh}.

It is also possible to consider a third type of flow, in which acoustic 
waves are favored over vortical movements. This third possibility to describe 
compressible turbulence, which we call compressible wave turbulence (CW), 
is dominated by acoustic waves and longitudinal velocity fluctuations.

\section{\label{Turb_hidro}Hydrodynamic equations}

We consider mass balance (continuity equation) and momentum balance 
(Navier-Stokes equations) for a compressible flow, with a polytropic 
state equation. In dimensionless units the equations are:

\begin{equation} \label{ec-NS-d}
    \frac{\partial \rho}{\partial t}= -\boldsymbol{\nabla} \cdot  (\rho \textbf{u})
\end{equation}

\begin{equation} \label{ec-NS}
    \begin{split} 
    \frac{\partial \textbf{u}}{\partial t}=\textbf{u}\times \boldsymbol{\omega} -\frac{1}{2}\boldsymbol{\nabla}\left ( \textbf{u}.\textbf{u} + \frac{2}{M^{2}(\gamma -1)}\rho^{\gamma -1}\right )\\ 
    ...+\frac{1}{\rho}\nu  _{c}\nabla^{2}\textbf{u}+\frac{1}{\rho}\left ( \xi _{c}+\frac{1}{3}\nu _{c} \right )\boldsymbol{\nabla}(\boldsymbol{\nabla}\cdot \textbf{u})+\textbf{f}
    \end{split}
\end{equation}

\begin{equation}
p=\rho^{\gamma }
\end{equation}

Here the dimensional units are $t_0$ for time $t$, $u_0$ for velocity $u$, 
$\rho_0$ for density $\rho$ and $p_0$ for pressure $p$. 
Length scale unit $l_0 = u_0 t_0$ is used for gradients. 
Dimensionless viscosities $\nu_c = \xi_c$ are considered as inverse of Reynolds number $R_e = u_0 l_0 / \nu$, where $\nu$ is the viscosity. 
Also  $\boldsymbol{\omega} = \boldsymbol{\nabla }\times \textbf{u}$ 
is the vorticity and $\textbf{f}$ the field of external forces. 

In the polytropic state equation we assume $\gamma = 5/3$, which corresponds to the 
isentropic flow for an ideal gas \cite{Chandrasekhar}. 
Finally, the nominal Mach number is $M = u_0/C_s$, with $C_{s}$ the speed of sound.

\section{Energy spectrum, frequency spectrum and correlation functions}

The wave number associated with scales of size $l$ is defined as $k=\frac{2\pi}{l}$. We can transform the velocity field $\textbf{u}$ from the physical space to the Fourier space $\hat{\textbf{u}}_{k}$. Through this transformation it is possible to perform calculations on the energy of the system expressing the distribution of energy between the vortices in terms of their spectrum. The spectrum $E(k) = \frac{1}{2}\left | \hat{\textbf{u}}_{k} \right |^{2}$ gives us the energy associated with each wave number regardless of its spatial distribution. 

We also consider the wavenumber-frequency spectrum $E_{ij}(\textbf{k},\omega )$, defined as \cite{Lugones}:

\begin{equation} \label{ec-espectro-esp-tem}
    E_{ij}(\textbf{k},\omega )=\frac{1}{2}\hat{u}^{*}_{i}(\textbf{k},\omega )\hat{u}_{j}(\textbf{k},\omega )
\end{equation}

where $\hat{u}_{i}(\textbf{k},\omega )$ is the Fourier transform in time and space of the i-component of the velocity field $\textbf{u}(\textbf{x},t )$, and where the asterisk indicates the complex conjugate.

The information of the time scales for each spatial mode and decorrelation times can be obtained from the temporal 
correlation function, given by the following relation \cite{Batchelor}:

\begin{equation} \label{ec-tiempo-corr}
    \Gamma _{ij}(\textbf{k},\tau )=\frac{\left \langle \hat{u}^{*}_{i}(\textbf{k},t)\hat{u}_{j}(\textbf{k},t+\tau ) \right \rangle_{t}}{ \left \langle \left | \hat{u}^{*}_{i}(\textbf{k},t)\hat{u}_{j}(\textbf{k},t )\right | \right \rangle_{t} }    
\end{equation}

where $\hat{u}_{i}(\textbf{k},t)$ is the Fourier transform of the i-component of the velocity field, the brackets indicate the time average, and only the real part is used. The correlation time $\tau _{D}(\textbf{k})$ is defined as the time at which the $\Gamma $ function falls to $1/e$ from its initial value. 

Although this definition is arbitrary, there are no quantitative differences between these different definitions except for a multiplicative factor of order one in the values of the correlation times.

\section{Characteristic times}

From the equations of motion and arguments of scale, it is possible to estimate different characteristic times of interest for our system. The turnover time of an eddy can be defined as $\tau _{nl}\sim \left [ ku(k) \right ]^{-1}$ where $k$ is the wave number and $u(k)$ is the amplitude of the velocity due to fluctuations at the scale $\sim 1/k$. It is the typical time in which a structure of size $\sim l$ suffers significant distortion due to the relative movement of its components. For the velocity scale $u\sim u_{rms}\left ( kL \right )^{-1/3}$, the nonlinear time scales in the inertial range, for a Kolmogorov prediction, can be written approximately as:

\begin{equation}
    t_{nl}=C_{nl}\left [ u_{rms}L^{-1/3}\left ( \sqrt{k_\perp ^{2}+k_{\parallel }^{2}} \right )^{2/3} \right ]^{-1}
\end{equation}

Where $C_{nl}$ is a dimensionless constant of unit order and $u_{rms}$ is defined as:

\begin{equation}
    u_{rms}=\left \langle \left | \textbf{u} \right |^{2} \right \rangle^{1/2}
\end{equation}

It is reasonable then that $\tau_{nl}$ is also the typical time of energy transfer of scales $\sim l$ to smaller scales.

The physics of the temporal correlation depends on other time scales, such as the characteristic sweep time on the scale $\sim 1/k$, which can be expressed as:

\begin{equation}
    t_{sw} = C_{sw}\left ( u_{rms}\sqrt{k_\perp ^{2}+k_{\parallel }^{2}} \right )^{-1}
\end{equation}

This time corresponds to the advection of small-scale structures by large-scale flow. Where $C_{sw}$ is another dimensionless constant of unit order.

Finally, the non-linear time of large scales $\tau_{NLG} $ independent of $k$, which should be distinguished from the non-linear time $\tau_{nl} $, will be used to estimate the time scale of energy transfer between scales. This time is defined as:

\begin{equation}
    \tau_{NLG} = \frac{l_{f}}{u_{rms}}
\end{equation}

where $l_{f}$ is the energy injection scale through the forcing.

\section{Theoretical Background}

In this section, we present the theoretical background of turbulent compressive flows of low Mach number with respect to the formation of sound waves and approximations to near incompressibility.
\vspace{4mm}

Klainerman and Majda postulate a set of dynamic equations that allows to describe flows of low Mach number. These equations are formed by incompressible hydrodynamic equations, a non-propagational density fluctuation field and other velocity and density fluctuations that obey acoustic equations. Since the dominant part of the asymptotic solution as the Mach number tends to zero, under a specific order of deficit and velocity fluctuations, are identical to the solutions of the pure incompressible equations, these equations are called: ''nearly incompressible'' (NI) hydrodynamics equations.

To obtain the solutions of nearly incompressible flow, the equations of compressible flows must be expanded in powers of Mach number ($M$). In this way we can separate the slow convective movements from the acoustics, treating the convective quantities as $\mathcal{O}(1)$. A condition that arises from this model is that the pressure variations must be of order $\mathcal{O}(M^{2})$, this is due to the longitudinal velocity fluctuations $\left ( \boldsymbol{\nabla} \cdot  \textbf{u}=\mathcal{O}(1) \right )$ and density variations $\left ( \delta  \rho = \mathcal{O}(M) \right )$ would violate the required pressure scale.

Then, in the NI model the fluctuations, in units of convective velocity $\left | \textbf{u} \right |= \mathcal{O}(1)$, must be $U_{T} \equiv \left | \boldsymbol{\nabla} \times \textbf{u} \right |= \mathcal{O}(1)$ and $U_{L} \equiv \left | \boldsymbol{\nabla} \cdot \textbf{u} \right | = \mathcal{O}(M^{2})$. Also, due tothe polytropic relationship, a constant average density $\rho = \mathcal{O}(1)$ is required, and the fluctuations must be $\delta \rho = \mathcal{O}(M^{2})$.
\vspace{4mm}

The NI model depends on assuming an almost incompressible medium. If we relax this condition and allows $\boldsymbol{\nabla}\cdot \textbf{u} = \mathcal{O}(1)$, then the medium can allow the formation of acoustic waves and vorticity movements with the same intensity. This situation is addressed by the Kraichnan \cite{MEC} model where $U_{L}$ and $U_{T}$ are both of order $\mathcal{O}(1)$ and the density fluctuations of order $\mathcal{O}(M)$. We will refer to this type of turbulence as modally equipartitioned compressible (MEC) turbulence.
\vspace{4mm}

Since even with very low Mach numbers a flow with zero vorticity can be initiated, i.e. pure longitudinal acoustic movements, 
it is possible to consider a third possibility to describe the compressible turbulence. In this third possibility, 
which we call compressible wave turbulence (CW), acoustic waves are favored over vorticity movements. 
This is an alternative to the nearly incompressible model (NI), which favors vorticity movements, or to the modally equipartitioned 
compressible (MEC) turbulence model, which treats the two types of movements with the same intensity.

For this model there are no theoretical basis on the scaling of $U_{L}$ and $U_{T}$ nor for $\delta \rho$. However, previous computational work by Ghosh and Matthaeus \cite{Ghosh} in 2-D domains of $64\times 64$ showed that $\delta \rho = \mathcal{O}(M)$. This order suggests that the CW turbulence obeys the MEC turbulence density scales. On the other hand for most of the reported simulation it was observed that $U_{L}$ is above $U_{T}$ suggesting that a smaller amount of vortical energy has been created.

\section{Numerical simulations}

We carried out 15 direct numerical simulations runs of the compressible fluid equations in turbulent regimes, in periodic cubic domains with a linear spatial resolution of $256^3$ grid points. The equations \ref{ec-NS-d} and \ref{ec-NS} were solved using a pseudospectral method, and evolved in time with a second order Runge-Kutta scheme.

A random forcing was used. This forcing has a parameter that allows us to determine its compressibility, so we can go from a pure incompressible forced system to a pure compressible forcing controlling this factor. The forcing is given by the equation:

\begin{equation} \label{forcing-eq}
    \textbf{F}_{\textbf{k}} = i\textbf{k}\times \hat{z}\psi _{1}-i\textbf{k}\psi _{2}
\end{equation}

where

\begin{equation} \label{Parametro_forzado}
    \begin{matrix}
    \psi _{1} = (1-a)f_{1}(\textbf{k})\\ 
    \psi _{2} = af_{2}(\textbf{k})
    \end{matrix}
\end{equation}

Here  $f_{i}(\textbf{k})$ is a function that is $1$ if $\textbf{k}$ is within the forced wavenumber range and is $0$ out of that range.

By modifying the value of the parameter $a$ the fluid can be compressibly excited $\left ( a=1 \right )$, generating thrusts parallel to the direction of movement and, alternatively, the incompressible modes are generated using $\left ( a=0 \right )$ pushing in the direction perpendicular to the direction of fluid movement. In addition, a temporal correlation is introduced that allows us to adjust the correlation time of the forcing.
The forcing is applied for wave numbers between $k_{min}=0.9$ and $k_{max}=1.8$.

For the runs, random initial conditions were used for the velocity field and null density fluctuations in the whole space. A random forcing described by the equation \ref{forcing-eq} was used. By means of this forcing we carry out three groups of simulations: one with purely compressible forcing, another with purely incompressible forcing and finally a mixed forcing that injects energy in the same amount to compressible and incompressible modes. In addition, for each of these three groups, five simulations were performed for different values of the Mach number.

The amplitude of the force remained almost constant for all simulations ($f_{o} \sim 0.1$) so that the energy converges to values of order $\mathcal{O}(1)$. The correlation frequency of the forcing (inverse of the correlation time) was set to $0.5$.

The viscosity of the system (inverse of the nominal Reynolds number) was set to $\nu_c = \xi_{c }=0.001$ in such a way that the dissipation scale is within the resolution range of all the simulations. The actual Reynolds number depends on the rms value of the velocity for each run.

The values of the Mach number used are in the range $\left [ 0.15,\; 0.55 \right ]$.

In the table \ref{valores-tabla}  the parameters of the simulations are detailed.

\begin{table}[h]
\centering
\begin{tabular}{|c||c|c|c|c|c|c|}
\hline
\textbf{run} & $R_{e}$ & $\epsilon$ & $k_{\eta}$ & M    & $\tau_{NLG}$ & $\bigtriangleup T_{nl}$ \\ \hline
\hline
\textbf{I1}                & 528       & 0.015          & 62          & 0.15 & 1.9            & 42                      \\ \hline
\textbf{I2}                & 522       & 0.015          & 62          & 0.25 & 1.9            & 42                      \\ \hline
\textbf{I3}                & 535       & 0.016          & 63          & 0.35 & 1.9            & 43                      \\ \hline
\textbf{I4}                & 533       & 0.015          & 62          & 0.45 & 1.9            & 43                      \\ \hline
\textbf{I5}                & 521       & 0.014          & 61          & 0.55 & 1.9            & 42                      \\ \hline
\hline
\textbf{M1}                & 408       & 0.006          & 49          & 0.15 & 2.5            & 33                      \\ \hline
\textbf{M2}                & 399       & 0.005          & 45          & 0.25 & 2.7            & 30                      \\ \hline
\textbf{M3}                & 397       & 0.005          & 47          & 0.35 & 2.5            & 32                      \\ \hline
\textbf{M4}                & 397       & 0.004          & 45          & 0.45 & 2.5            & 32                      \\ \hline
\textbf{M5}                & 420       & 0.005          & 47          & 0.55 & 2.4            & 34                      \\ \hline
\hline
\textbf{C1}                & 268       & 3.6E-5          & 14          & 0.15 & 3.7            & 21                      \\ \hline
\textbf{C2}                & 330       & 5.2E-5          & 15          & 0.25 & 3.0            & 26                      \\ \hline
\textbf{C3}                & 329       & 1.2E-4          & 18          & 0.35 & 3.0            & 26                      \\ \hline
\textbf{C4}                & 280       & 8.3E-5          & 17          & 0.45 & 3.6            & 22                      \\ \hline
\textbf{C5}                & 189       & 3.0E-5          & 13          & 0.55 & 5.3            & 15                      \\ \hline
\end{tabular}
\caption{\footnotesize {Simulation data with $256^{3}$ grid points. The notation I, M, and C refer to the type of forcing (I $=$ incompressible forcing - M $=$ mixed forcing - C $=$ compressible forcing). The average Reynolds number value ($R_{e}$) that the simulation reaches in its steady state is indicated. It also indicated the dissipation rate $\epsilon$ and dissipation scale $k_{\eta}$. The value of the Mach number (M) used in each simulation is specified. $\bigtriangleup T_{nl}=T/\tau_{NLG}$ is the total number of non-linear times considered in each run.}}
\label{valores-tabla}
\end{table}

\section{Results}

\begin{figure}[H]
\centerline{
  \includegraphics[width=0.4\textwidth]{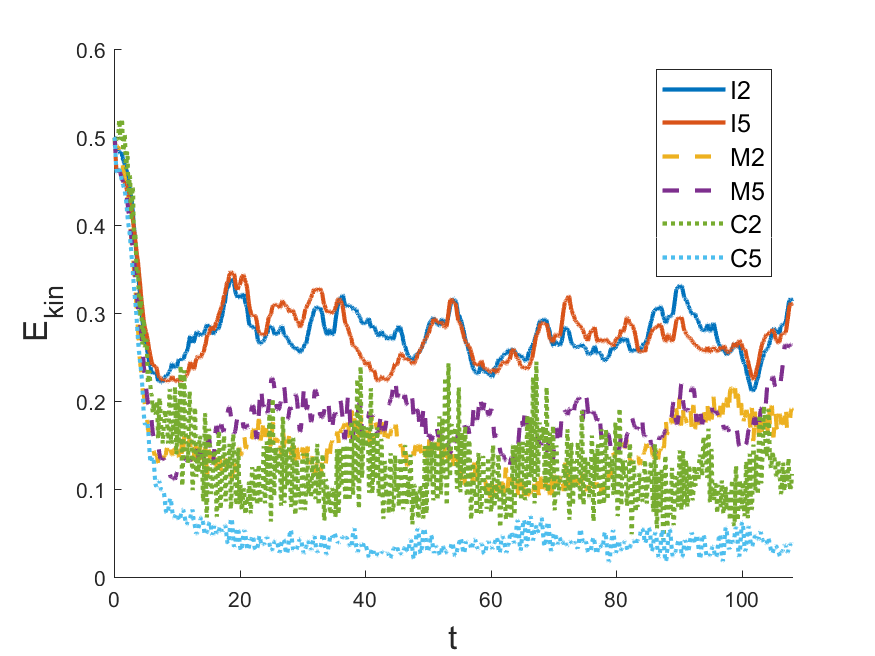}}
  \caption{\footnotesize {Kinetic energy as a function of time (simulations of the table \ref{valores-tabla}).}}
  \label{fig:energia_cinetica}
\end{figure}

The figure \ref{fig:energia_cinetica} shows the evolution of kinetic energy, defined as $E_{c}(t)= \left \langle \left | \textbf{u} \right |^{2} \right \rangle_{\textbf{x}}$, for the simulations specified in the table \ref{valores-tabla}.

The energy decays initially until $t \sim 20$, when a statistically steady state is reached with 
$E_{c} \sim 0.2$. 
At this point, the cascade of energy excite the dissipation scales and the system begins to dissipate energy significantly.

We identify strong oscillations of the energy when the system reaches the steady state. This effect is attributed to the development of waves in the fluid, and as can be seen, this effect is even greater for high Mach values and compressible forcing. Instead, for incompressible forcing, the kinetic energy has a smoother profile and it does not change when modifying the value of the Mach number. This allows us to infer, through an early analysis, the existence of waves in the fluid and that this is strongly dependent on the Mach number and the applied force.

\begin{figure}[H]
\centerline{
  \includegraphics[width=0.4\textwidth]{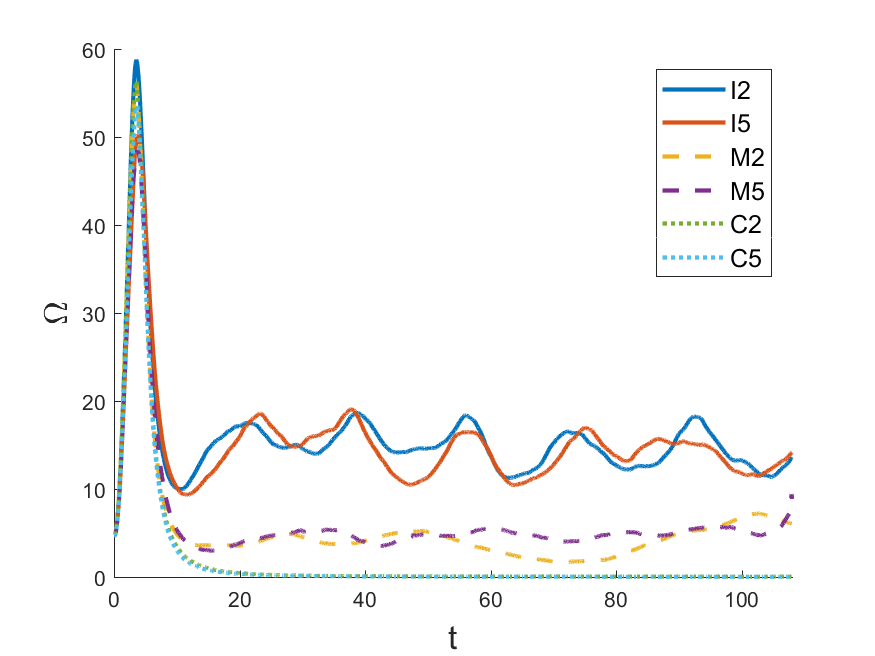}}
  \caption{\footnotesize {Enstrophy as a function of time (simulations of the table \ref{valores-tabla}).}}
  \label{fig:enstrofia}
\end{figure}

To determine the stationary range, the system was allowed to evolve until the enstrophy is stabilized around an asymptotic value (see Fig: \ref{fig:enstrofia}).

We can observe in the simulations $C_{i}$ (compressible forcing) that the enstrophy is practically null when the steady state is reached. 

Analysis of spatio-temporal spectra and correlation times were made in a time window $I_{t} = \left [ t_{o},t_{f} \right ]=\left [ 28,108 \right ]$. This corresponds to between $15$ and $43$ non-linear times.

\subsection{Spectral analysis} \label{Análisis espectral}

Before proceeding to the analysis of the wave number and frequency spectrum, and to the study of the correlation time for each mode, we present the spectra in the space of wave numbers, as is usually done to characterize the turbulent flows. In addition to being useful for characterizing simulations, these spectra will also be important to identify the behavior of different modes depending on what dynamical times are expected to be dominant.

We analyze the energy spectra $E(k)$ for the different types of forcing. This spectrum is compared with the law of scales, $k^{-5/3} $, predicted by the Kolmogorov theory for incompressible, isotropic and homogeneous turbulent flows \cite{Taylor1,Taylor2}. A fit is made on the spectrum $E(k)$ and the power law is determined in each case. To find the characteristic exponent, a linear adjustment of minimum squares is applied to the data in a logarithmic scale (restricted to the wave numbers belonging to the inertial range).

\begin{figure}[H]
\centering
\subfigure[I2]{\includegraphics[width=0.22\textwidth]{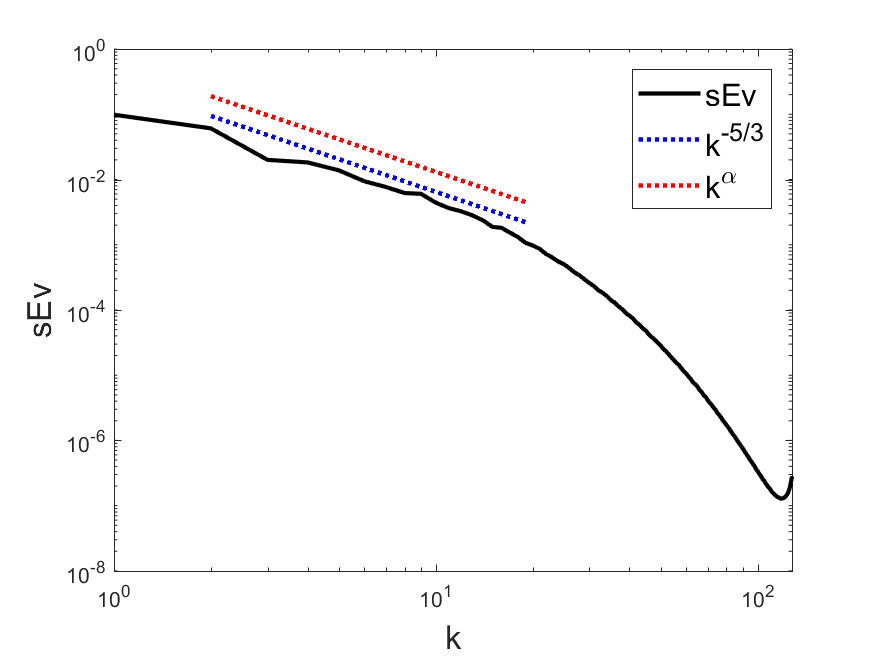}}
\subfigure[M2]{\includegraphics[width=0.22\textwidth]{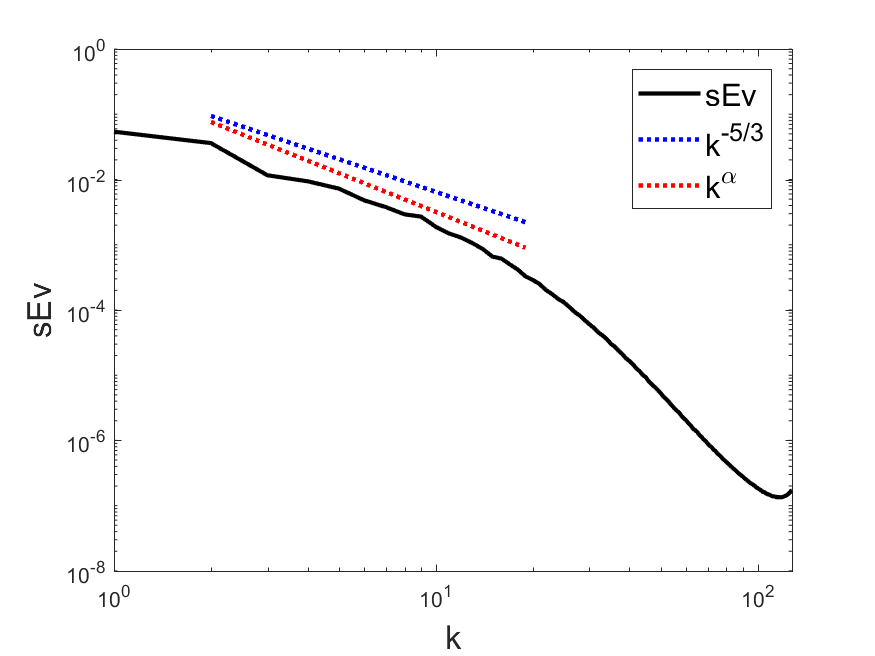}}
\subfigure[C2]{\includegraphics[width=0.22\textwidth]{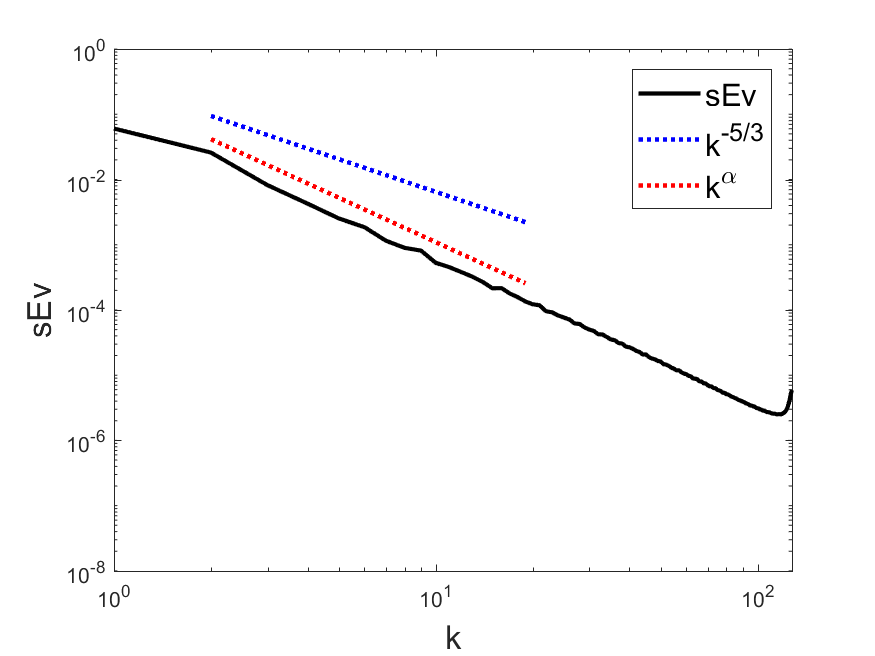}}
\caption{Energy spectra for simulations with Mach $0.25$ averaged in the time window $I_ {t}$. Above the spectra, the slopes corresponding to Kolmogorov law and those corresponding to the linear adjustment are shown.} \label{fig:espectrosA}
\end{figure}

In the figure \ref{fig:espectrosA} the energy spectra for the simulations with incompressible, mixed and compressible forcing, for a Mach value of $0.25$, are shown. 

Although the Kolmogorov law is only valid for incompressible flows, the simulations with low Mach number and incompressible forcing show an exponent $\alpha = -1.66 \pm 0.06$ (for the simulation I2) compatible with the prediction of Kolmogorov $k^{-5/3}$ for isotropic and homogeneous turbulence. In addition, the spectrum presents a similar profile for all Mach number values. The rest of the Mach number values used presented a similar behavior.

The linear adjustment for the compressive forcing simulation C2 returns a value of 
$\alpha  = -2.26 \pm 0.03$. These spectra present a more pronounced decay compared to Kolmogorov law. This should be expected since the energy cascade has an additional channel from eddies to acoustic modes that are then damped, as the model of Zank and Matthaeus predicts.

These spectra also show a greater preponderance of the large scales compared to the small scales, a result that is consistent with the low values of enstrophy reported for these simulations in the previous section and with results of structure formation that we will see later.

Also, all the spectra present a similar profile for all values of the Mach number. These simulations reflect that, at low Mach numbers, the power law of the spectra does not present a remarkable dependence with the value of the Mach number. This situation could become different for Mach values greater than unity, where we move away from nearly incompressibility.

\subsection{Axisymmetric spectrum}

To continue with the analysis of the energy spectral distribution, the level curves of the axisymmetric spectra of kinetic energy are presented below as a function of $k_{x}$ and $k_{y}$, defined by the equation:

\begin{equation}
    \textbf{e}(k_{x},k_{y }) =  \left \langle \hat{\textbf{u}}^{2}(k_{x},k_{y }) \right \rangle_{t}
\end{equation}

\begin{figure}[H]
\centering
\subfigure[I2]{\includegraphics[width=0.33\textwidth]{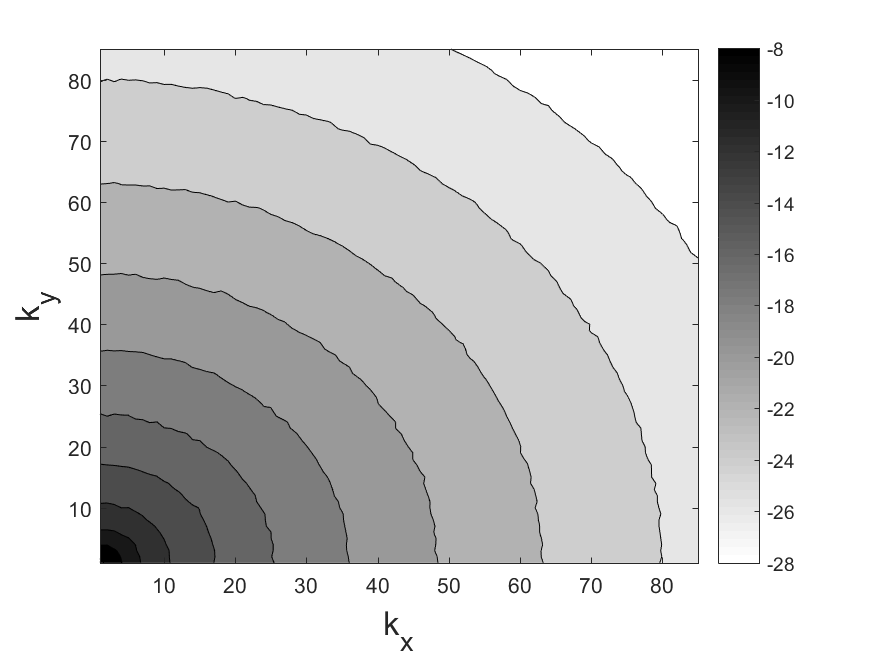}}
\subfigure[M2]{\includegraphics[width=0.33\textwidth]{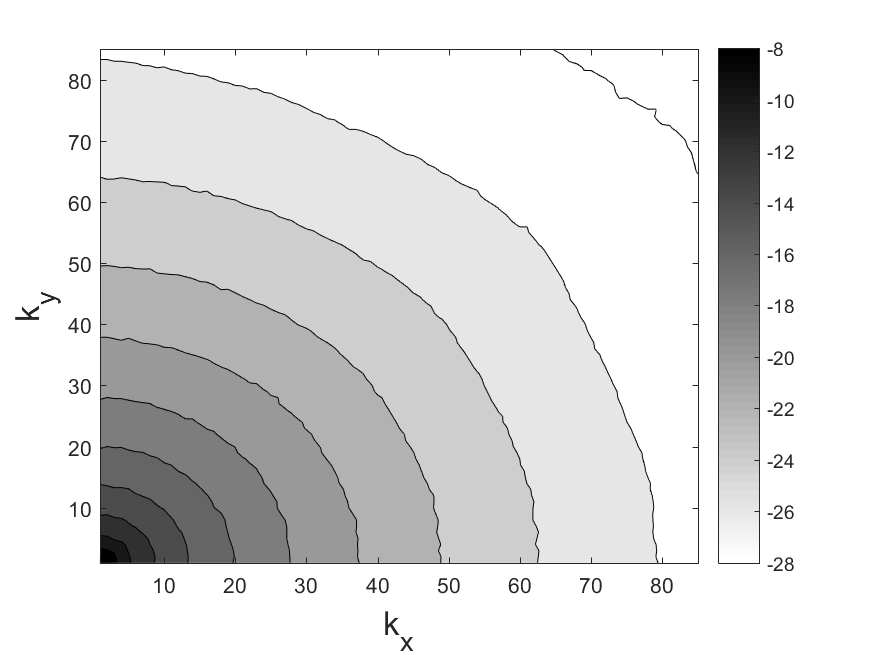}}
\subfigure[C2]{\includegraphics[width=0.33\textwidth]{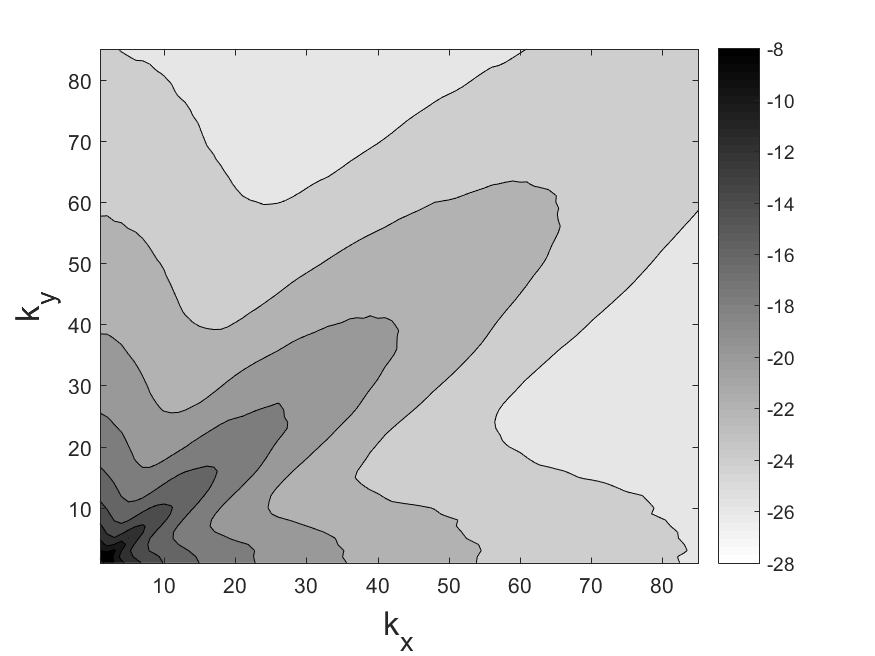}}
\caption{Axisymmetric spectra averaged over the time window $ \left [ t_{o},t_{f} \right ]=\left [ 28,108 \right ]$. The logarithmic grayscale indicates the spectral amplitude at each point of the plane $k_{x}$, $k_{y}$.} 
\label{fig:axisimetA}
\end{figure}

In figure \ref{fig:axisimetA} the axisymmetric spectra are presented for the simulations with Mach $0.25$ and for the different types of forcing.

The logarithmic grayscale indicates the intensity of $\textbf{e}(k_{x},k_{y})$ in each point of the Fourier space.

For the I2 simulation it is possible to observe a semicircle profile of the contour lines, consistent with previous simulations for incompressible, isotropic and homogeneous flows.

On the other hand, for the simulations C2, and to a lesser extent for M2, the spectra show higher energy near the points with $k_{x}=k_{y}$. Especially for Mach values less than $0.3$ the axisymmetric spectra show a remarkable deformation around the line $k_{x}=k_{y}$.

These effects have not been observed in simulations or previous experiments under conditions of near incompressibility. The reason for the increase in energy in the axisymmetric spectra near the points with $k_{x}=k_{y}$ requires a more detailed analysis that we plan to study in future work.

This behavior is then reflected in the formation of a number of significant large-scale structures, absent in the rest of the runs.

\begin{figure}[H]
\centering
\subfigure[C1]{\includegraphics[width=0.22\textwidth]{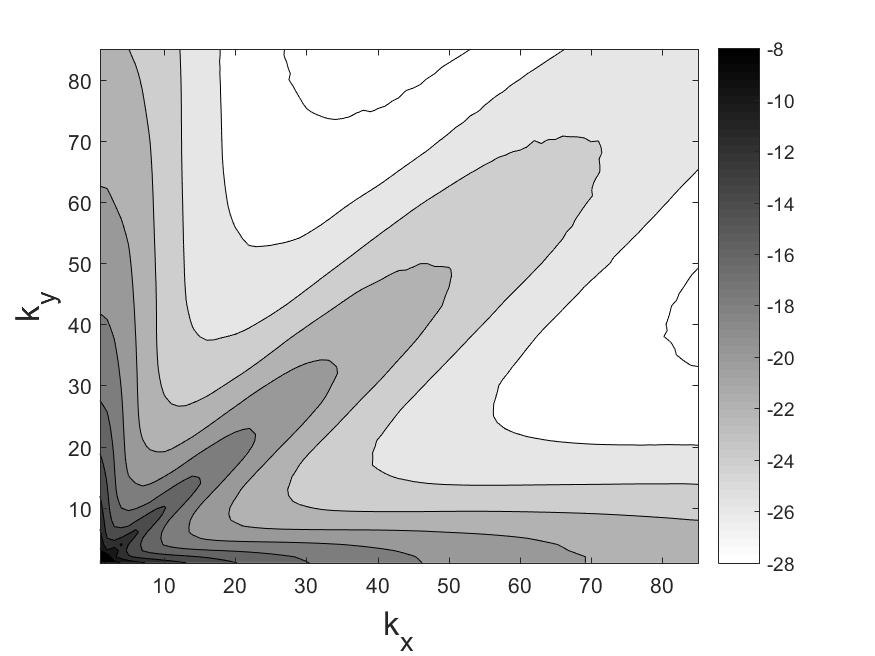}}
\subfigure[C2]{\includegraphics[width=0.22\textwidth]{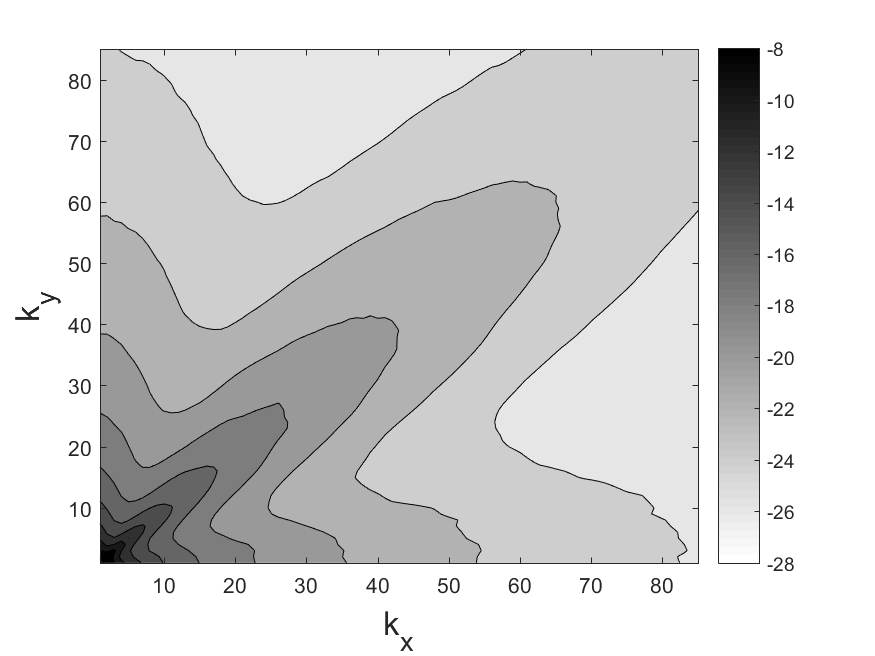}}
\subfigure[C3]{\includegraphics[width=0.22\textwidth]{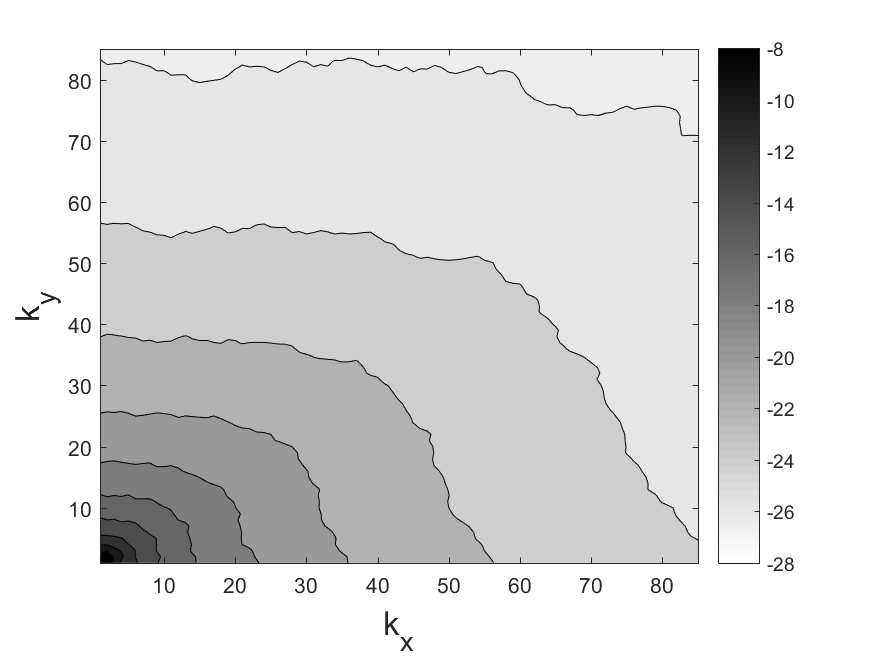}}
\subfigure[C4]{\includegraphics[width=0.22\textwidth]{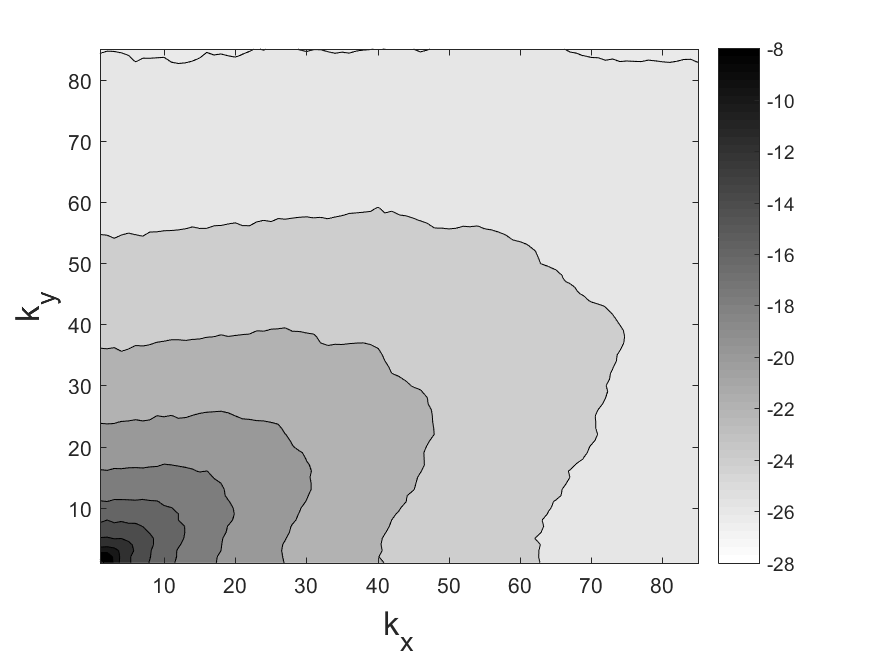}}
\subfigure[C5]{\includegraphics[width=0.22\textwidth]{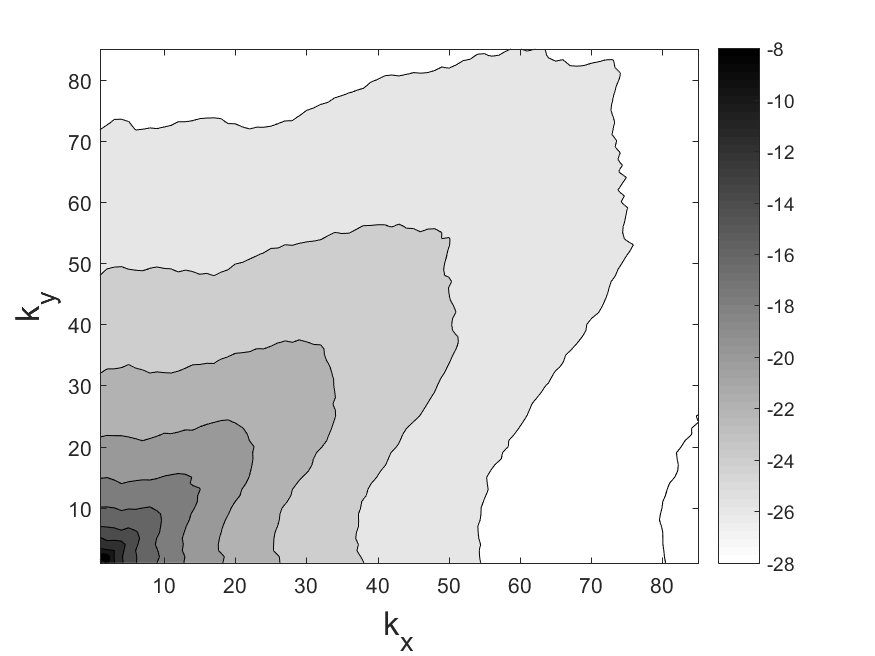}}
\caption{Axisimetric spectra averaged over the time window $ \left [ t_{o},t_{f} \right ]=\left [ 28,108 \right ]$ for simulations with compressible forcing. The logarithmic grayscale indicates the spectral amplitude at each point of the plane $k_{x}$, $k_{y}$.} \label{fig:axisimetC}
\end{figure}

Also in figure \ref{fig:axisimetC} for the axisymmetric spectra, we can see the effect of changing the Mach number for all the simulations with compressible forcing.

\subsection{Spatial-temporal spectra}

In this section we analyze the spatio-temporal spectra of the velocity fields $u_{x}$ and $u_{y}$, in order to detect the presence (or absence) of sound waves. To perform this type of analysis, the fields must be stored with very high frequency cadence in order to resolve the waves in time and space; 
in particular, $dt =5\cdot 10^{-3}$ was taken as the temporal sampling rate.

Since the spatio-temporal spectra depend on $\textbf{k}=(k_{x},k_{y},k_{z})$ and on the frequency $\omega$, to visualize the spectra more conveniently, some of the components of $\textbf{k}$ are fixed; also, the spectra are normalized for each wave number $k$.

\begin{figure}[H]
\centering
\subfigure[I2]{\includegraphics[width=0.33\textwidth]{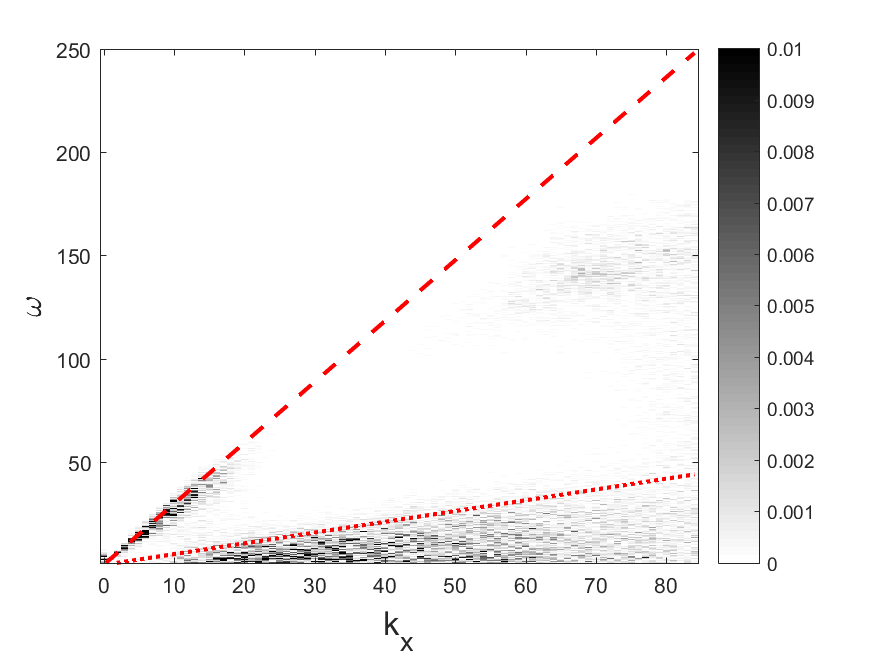}}
\subfigure[M2]{\includegraphics[width=0.33\textwidth]{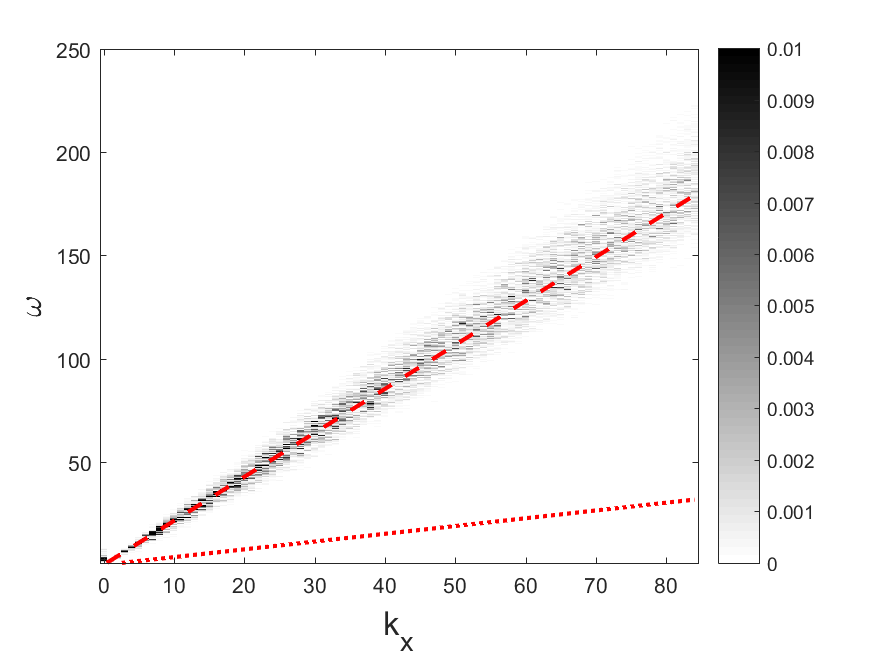}}
\subfigure[C2]{\includegraphics[width=0.33\textwidth]{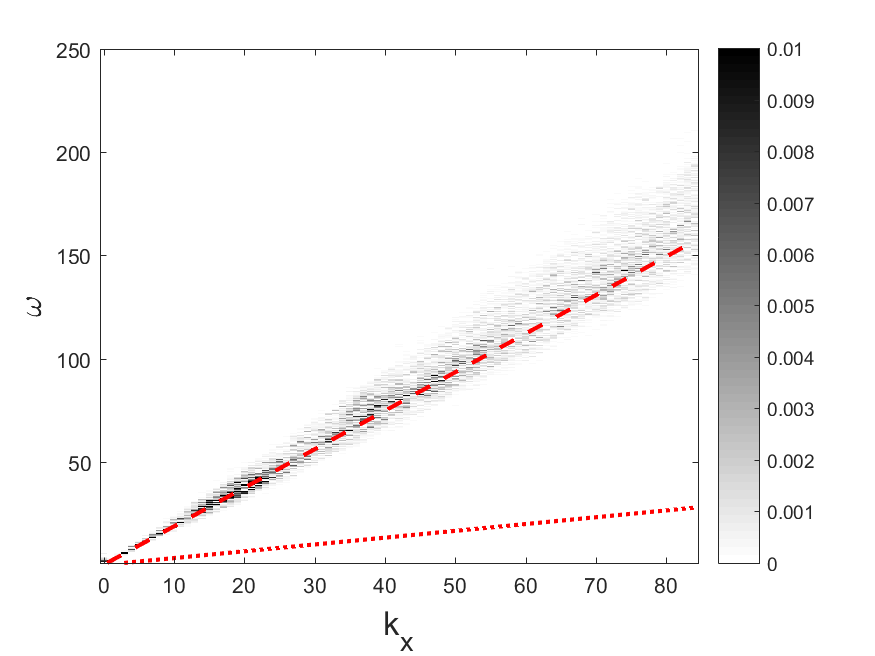}}
\caption{Spatio-temporal power spectrum $E_{xx}(k_{x},\omega )$. The line (--) shows the dispersion relation of sound waves, the line (:) is the sweeping relation.} \label{fig:espacAx}
\end{figure}

\begin{figure}[H]
\centering
\subfigure[I2]{\includegraphics[width=0.22\textwidth]{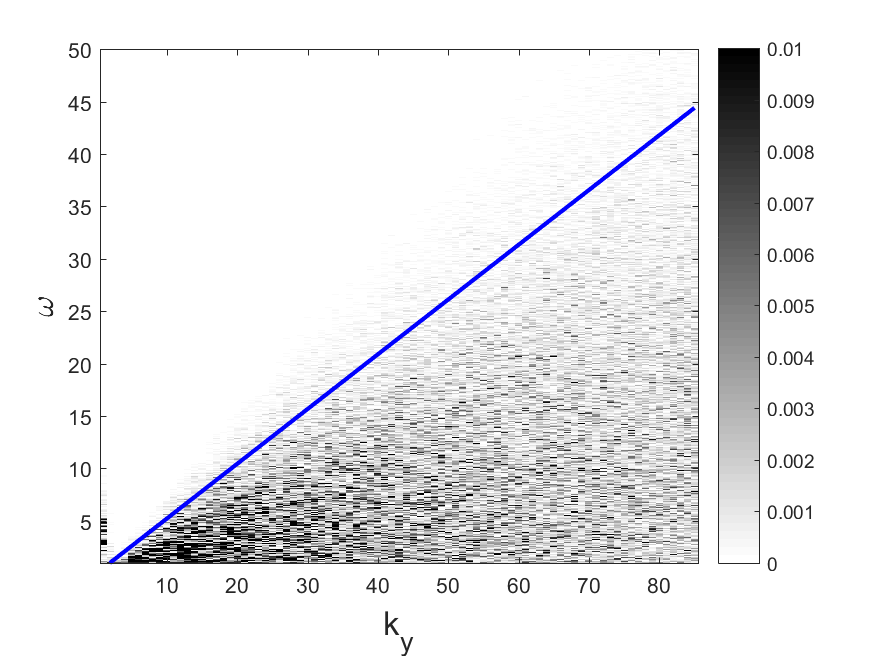}}
\subfigure[M3]{\includegraphics[width=0.22\textwidth]{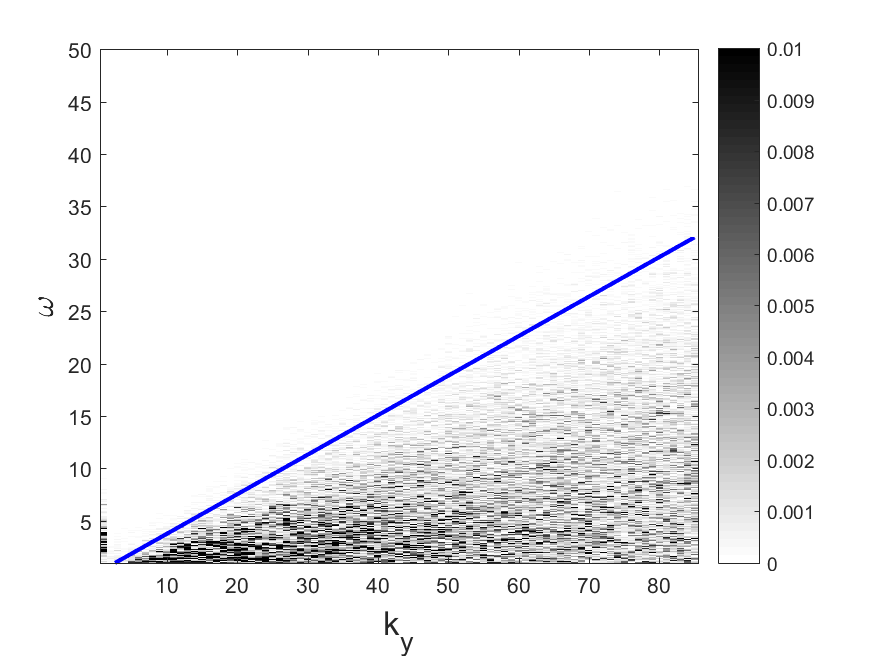}}
\subfigure[C2]{\includegraphics[width=0.22\textwidth]{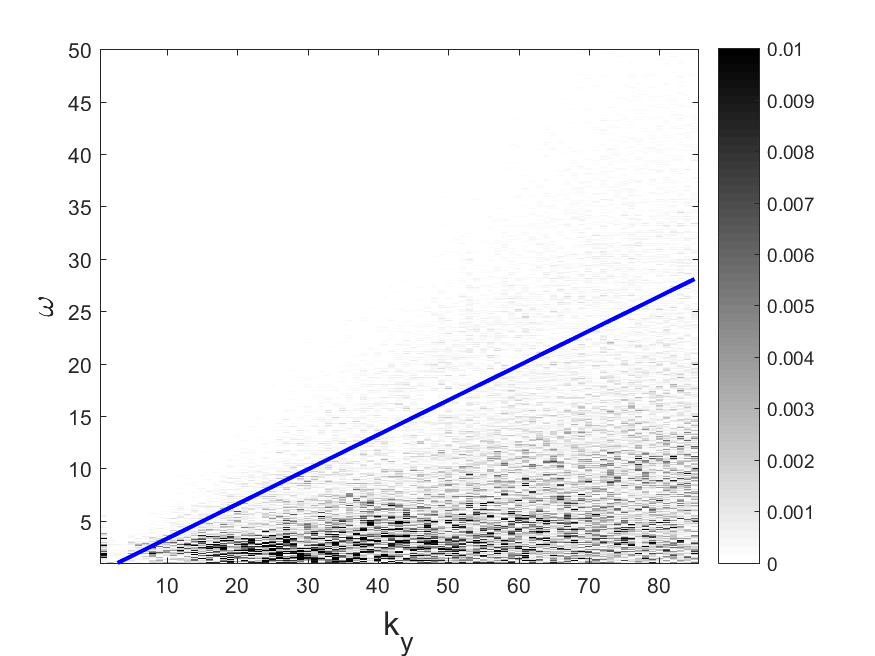}}
\caption{Space-time power spectrum $E_{xx}(k_{y},\omega )$. The straight line shows the sweeping relation.} \label{fig:espacAy}
\end{figure}

Figure \ref{fig:espacAx} shows the spectra $E_{xx}(k_{x},\omega )$ with $k_{y}=k_{z}=0$. In 
figure \ref{fig:espacAy} we show the spectra $E_{xx}(k_{y},\omega )$ with $k_{x}=k_{z}=0$ for 
Mach number $=0.25$ and the different forcing types.

The spectra $E_{xx}(k_{x},\omega )$, for incompressible forcing have higher energy around the dispersion relation of sound waves only for low $k$, in particular less than 20. For higher values of $k$ the energy is concentrated below the sweeping relation. That is, for simulations with incompressible forcing, for small $k$ the propagation of sound waves dominates, while for high $k$ turbulent effects are dominant. As the value of the Mach number is increased, the sound waves take on greater ''protagonism'' and the concentration of energy below the sweeping relation decreases, although it is maintained to some extent.

For compressible forcing, the spectra $E_{xx}(k_{x},\omega )$ have more energy around the dispersion relation of the sound and this is true for every value of $k$. In a similar way to the mixed forcing case, in this case even for high $k$ the spectra present higher energy around the dispersion relation of the sound. It is observed that for high $k$ the spectrum presents a wider profile, and tends to focus above the dispersion relation of the sound. In $E_{xx}(k_{x},\omega )$ energy is not observed around the sweeping relation. 

For the spectra $E_{xx}(k_{y},\omega )$, for all runs, the dispersion relation of the sound is absent (the propagation of the sound occurs only in the longitudinal component), so that the system is dominated by pure turbulence. The concentration of energy does not present intensity changes when modifying the Mach number in these spectra.

\subsection{Correlation times}

The relevant information for temporal scales can also be obtained from the temporal decorrelation function. In this section we present the correlation times described by the equation \ref{ec-tiempo-corr}. 
Similar to the previous section, to study the correlation times associated with sound waves we look at $\Gamma_{xx}(k_{x},\tau)$ with $k_{y}=k_{z}=0$. To study the rest of turbulence characteristic times we analyze the correlation times from $\Gamma_{xx}(k_{y},\tau)$ with $k_{x}=k_{z}=0$. Correlation times are defined as the time when the $\Gamma$ function decays to $1/e$ of its initial value.

\begin{figure}[H]
\centering
\subfigure[I2]{\includegraphics[width=0.33\textwidth]{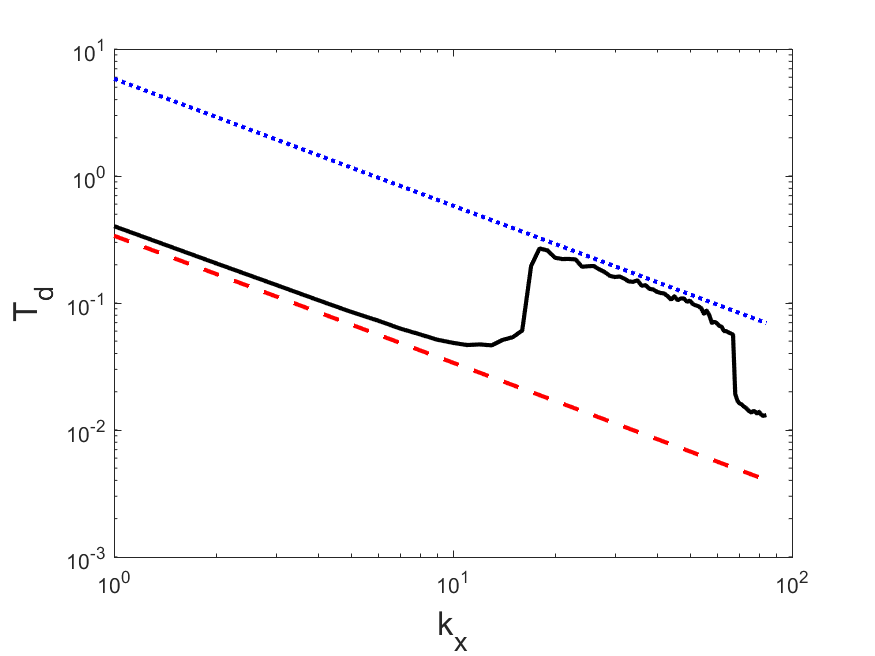}}
\subfigure[M2]{\includegraphics[width=0.33\textwidth]{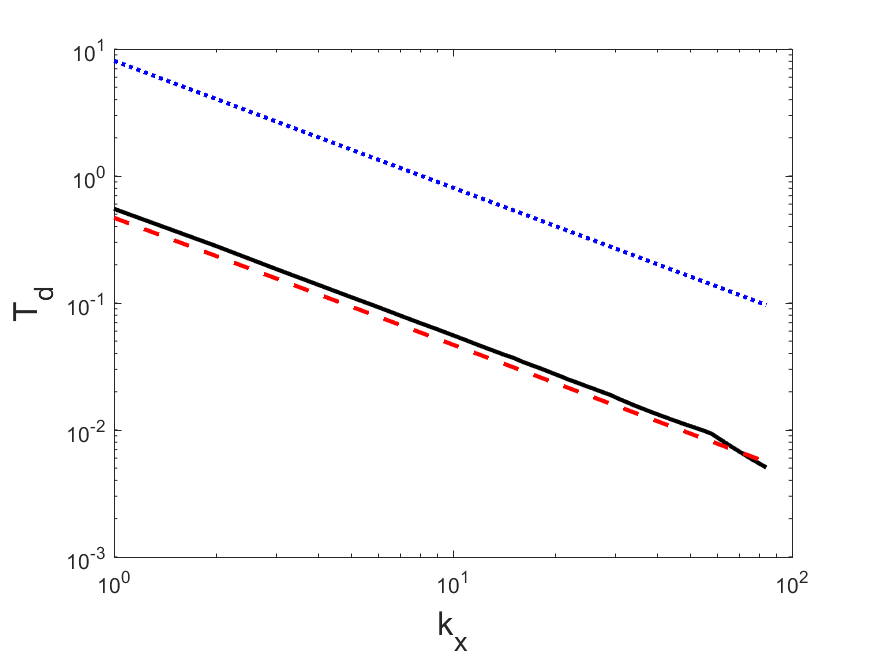}}
\subfigure[C2]{\includegraphics[width=0.33\textwidth]{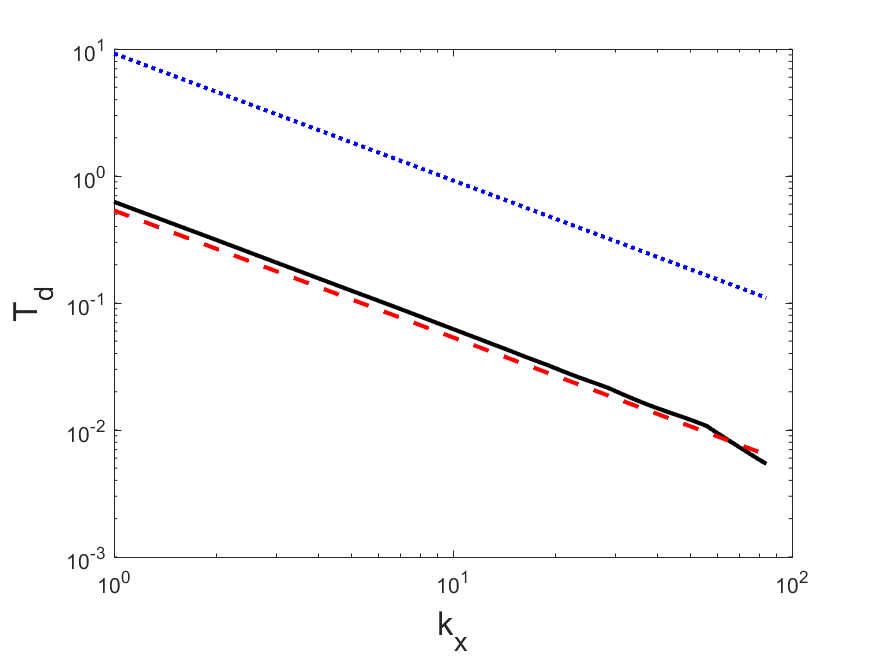}}
\caption{Correlation times $Td$ for $\Gamma_{xx}(k_{x},\tau)$. The curve (--) shows the characteristic time associated with sound waves, the line (:) is the characteristic time of sweeping.} \label{fig:correlcAx-1}
\end{figure}

\begin{figure}[H]
\centering
\subfigure[I2]{\includegraphics[width=0.22\textwidth]{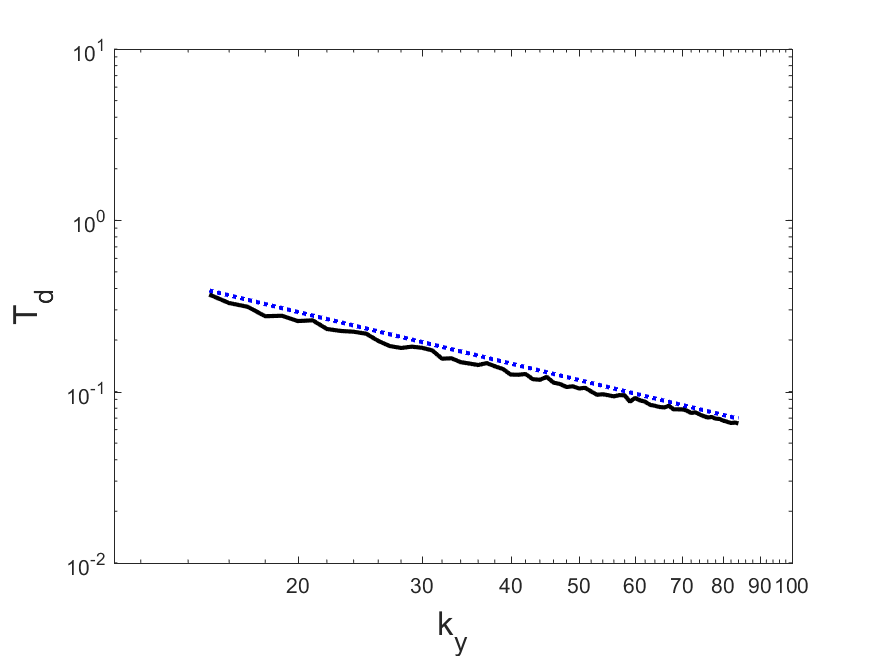}}
\subfigure[M2]{\includegraphics[width=0.22\textwidth]{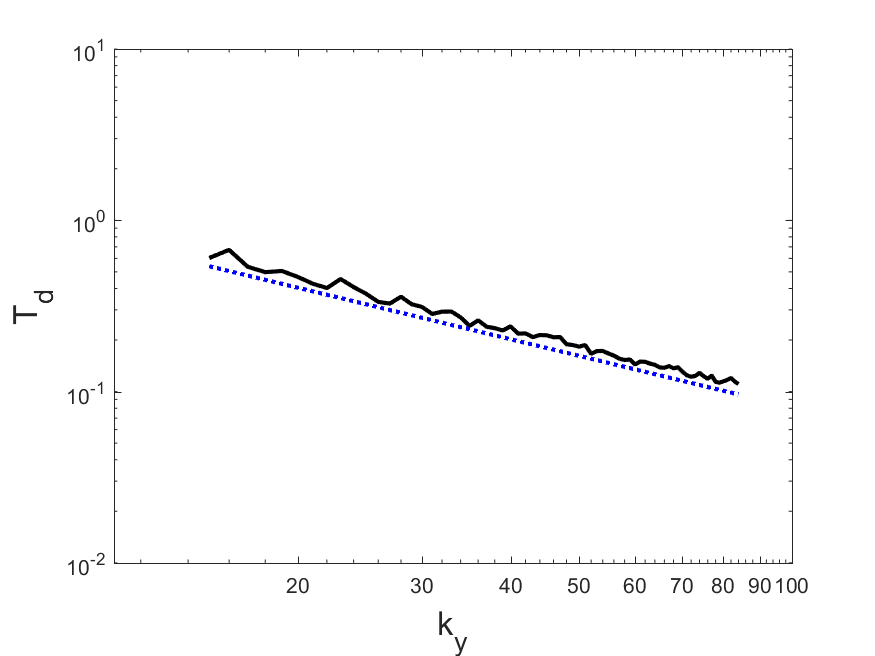}}
\subfigure[C2]{\includegraphics[width=0.22\textwidth]{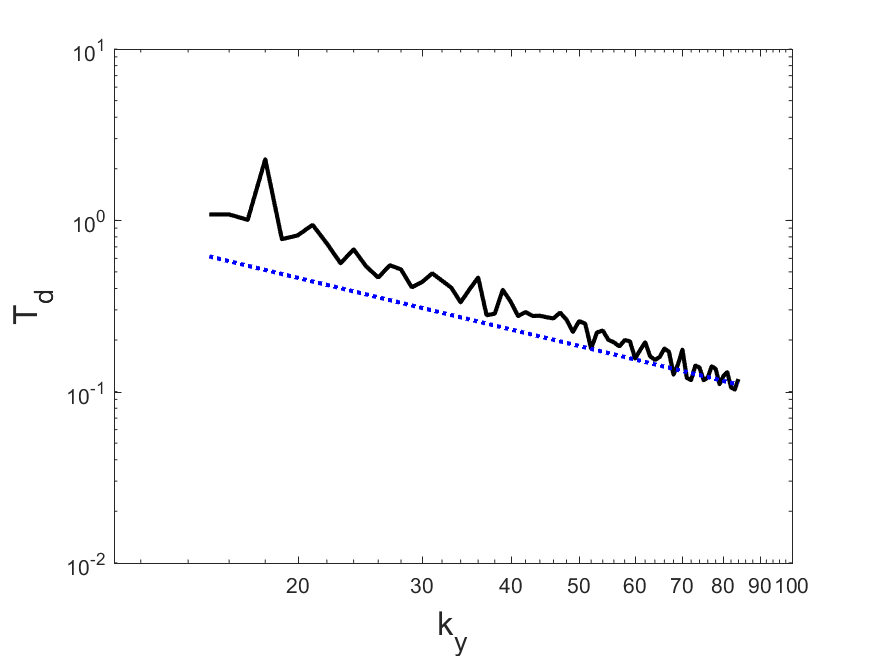}}
\caption{Correlation times $Td$ for $\Gamma_{xx}(k_{y},\tau)$. The curve (:) shows the characteristic time associated with the sweeping time.} \label{fig:correlcAy-1}
\end{figure}

In figure \ref{fig:correlcAx-1} we show the correlation times $Td_{x}(k_{x})$ with $k_{y}=k_{z}=0$. In figure \ref{fig:correlcAy-1} we show the characteristic times $Td_{x}(k_{y})$ with $k_{x}=k_{z}=0$ for Mach $0.25$ and the different forcing types.

In $Td_{x}(k_{x})$, for incompressible forcing, the correlation times show a jump. For low $k$ the system is dominated by the characteristic times associated with sound waves, while for high $k$ dominate the characteristic times associated with sweeping. This is consistent with the spatio-temporal spectra analyzed in the previous section, where a similar behavior was observed. For higher Mach values this effect is shielded and the characteristic times of the acoustic waves are dominant, even for large $k$ values. The values of $k$ for which this change in behavior occurs appear to be strongly correlated with the Mach number. We assume that this behavior could become an effect of turbulent structure formation.

For mixed and compressible forcing we don't observe the jump in the characteristic times reported for the incompressible forcing case.

\begin{figure}[H]
\centering
\subfigure[I1]{\includegraphics[width=0.22\textwidth]{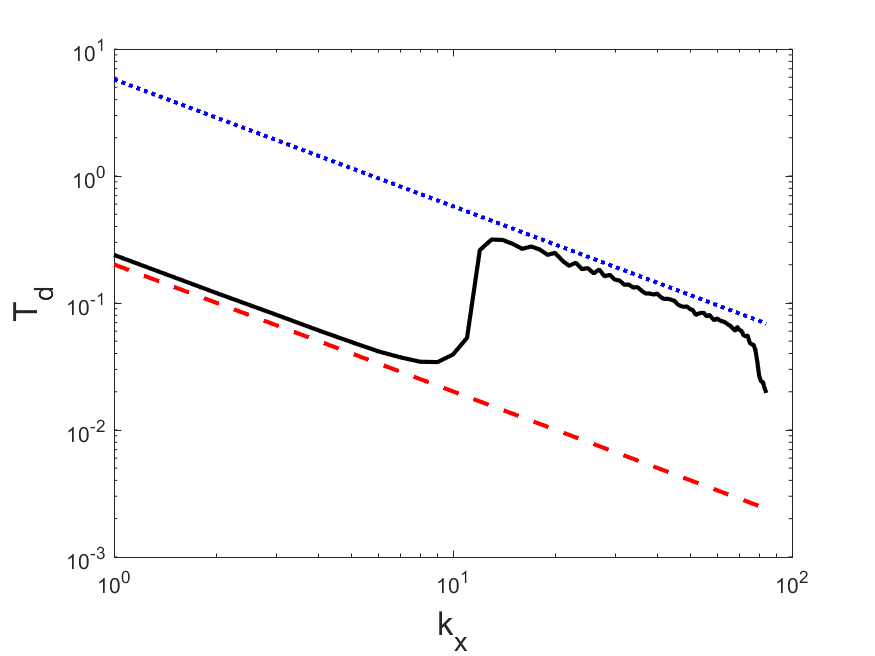}}
\subfigure[I2]{\includegraphics[width=0.22\textwidth]{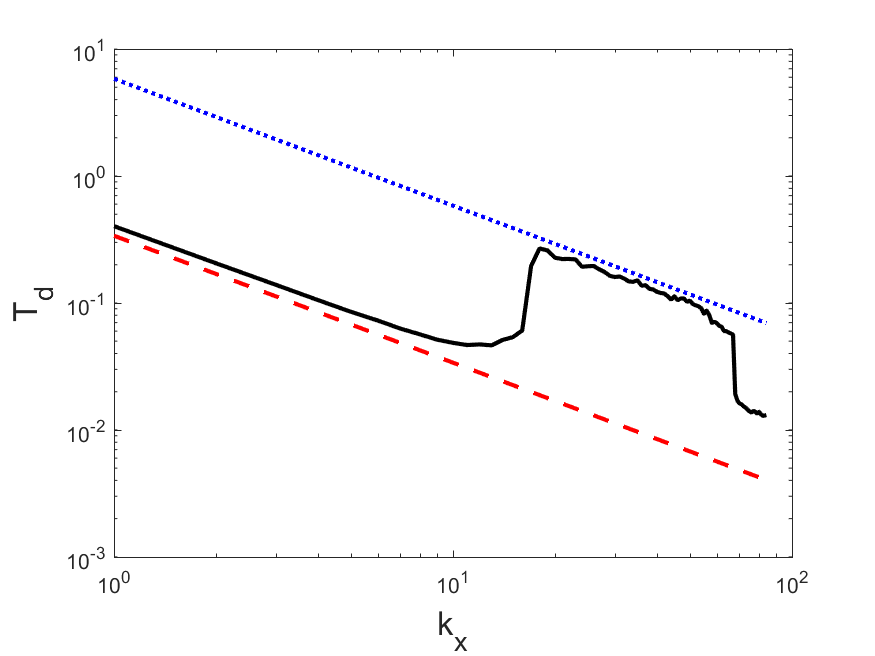}}
\subfigure[I3]{\includegraphics[width=0.22\textwidth]{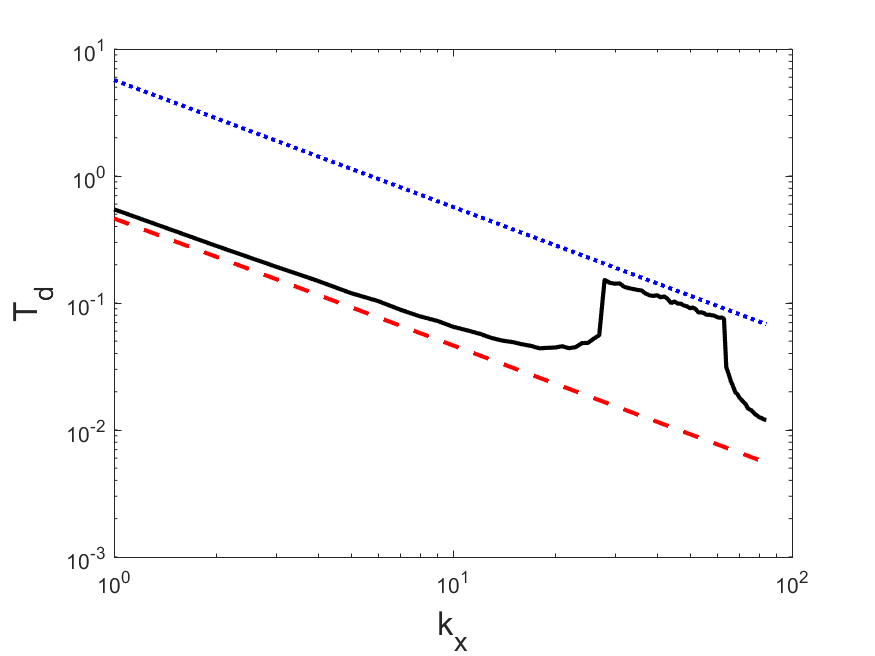}}
\subfigure[I4]{\includegraphics[width=0.22\textwidth]{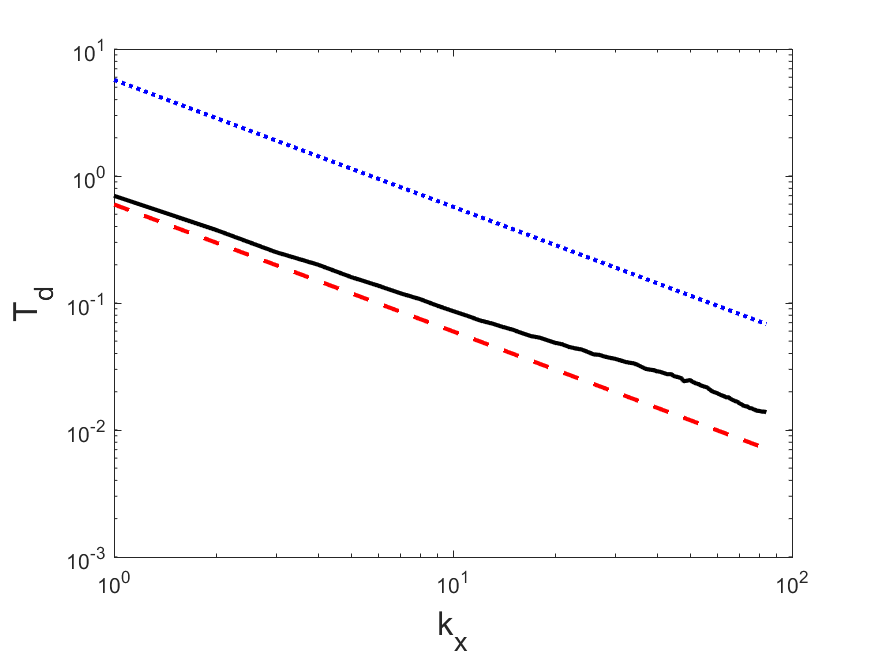}}
\subfigure[I5]{\includegraphics[width=0.22\textwidth]{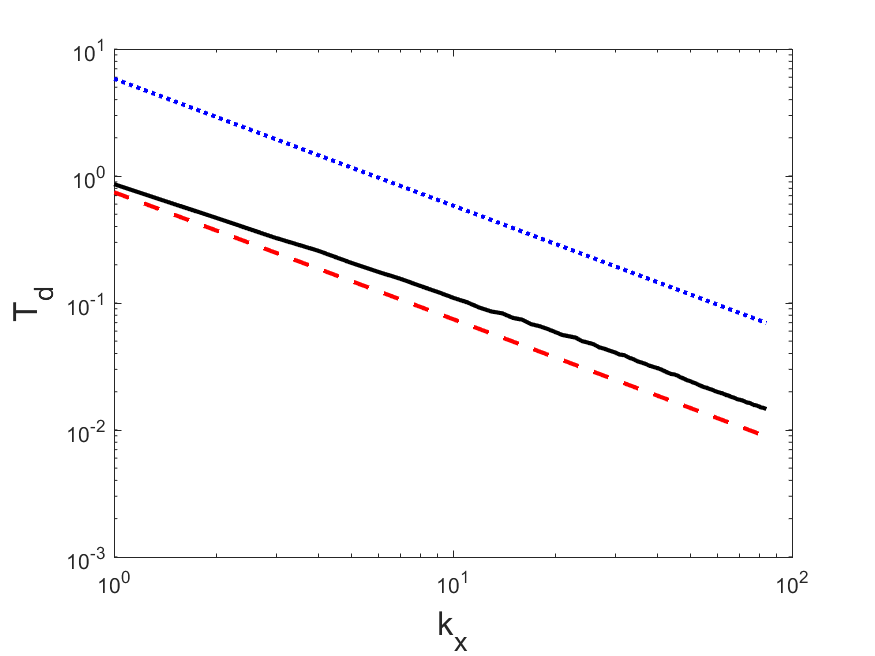}}
\caption{Correlation times $Td$ for $\Gamma_{xx}(k_{x},\tau)$ for simulations with incompressible forcing. The curve (--) shows the characteristic time associated with sound waves, the line (:) is the characteristic time of sweeping.} \label{fig:nncorrelcAx-1}
\end{figure}

Also we present in figure \ref{fig:nncorrelcAx-1} the correlation times for all the simulations with incompressible forcing and different Mach numbers.

\subsection{Density fluctuations}

To corroborate the predictions described by the NI, MEC and CW theories the density fluctuations and its dependence on the Mach number are calculated. Density fluctuations were calculated at different times (after the stabilization time $t_{o}$) and for different values of Mach number ($M = 0.15$ - $0.25$ - $0.35$ - $0.45$ - $0.55$).

The values shown, for each value of Mach number, consist of the average of the measurements at each time and the error bars to the dispersion of them.

In addition, an adjustment of type $\alpha M^{\beta}$ was made and the exponent $\beta$ was compared with that predicted by the different theories.

\begin{figure}[H]
\centerline{
  \includegraphics[width=0.4\textwidth]{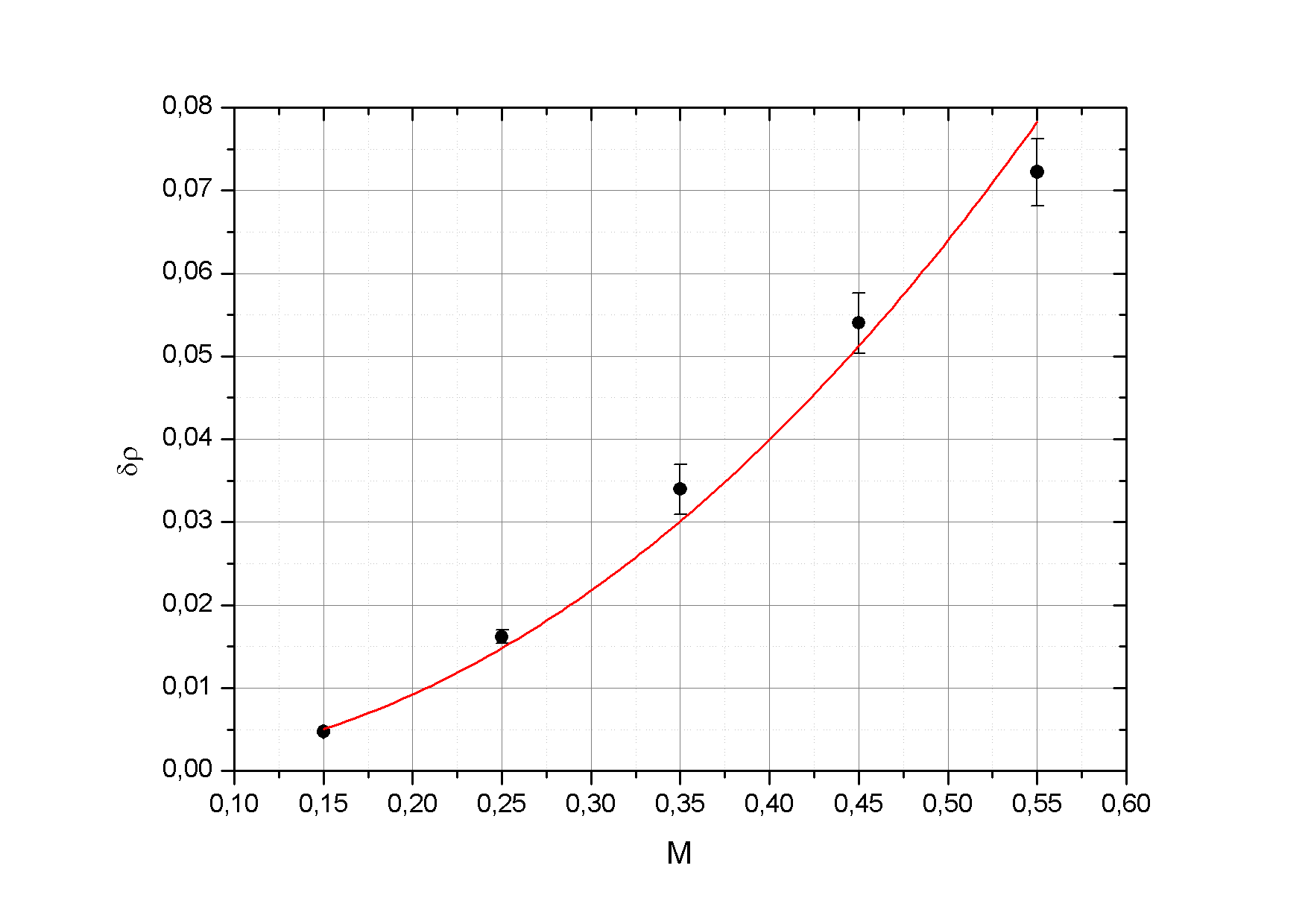}}
  \caption{\footnotesize {Density fluctuations vs Mach number, for incompressible force.}}
  \label{fig:densidad-SI1}
\end{figure}

\begin{figure}[H]
\centerline{
  \includegraphics[width=0.4\textwidth]{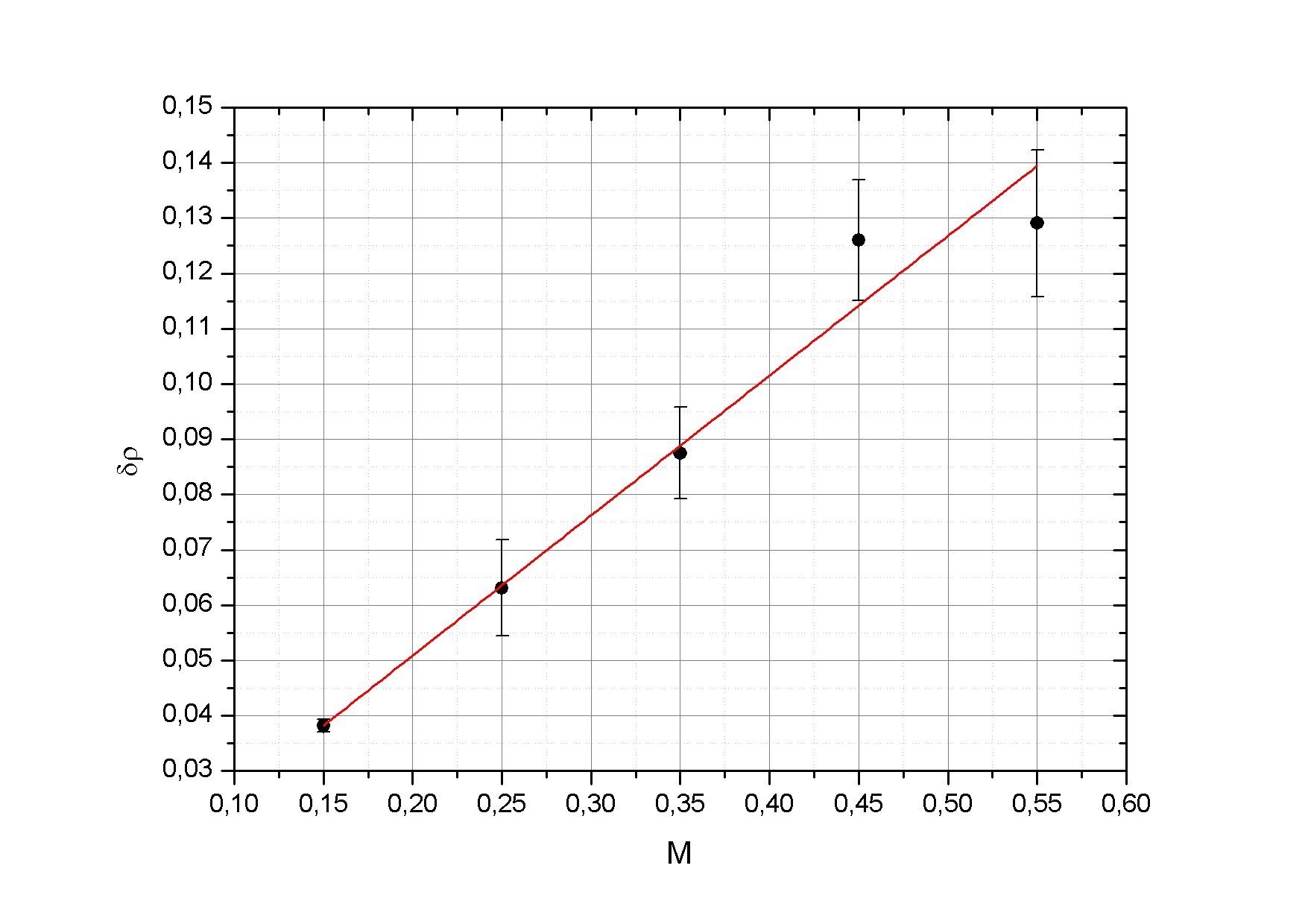}}
  \caption{\footnotesize {Density fluctuations vs Mach number, for mixed forcing.}}
  \label{fig:densidad-SI2}
\end{figure}

\begin{figure}[H]
\centerline{
  \includegraphics[width=0.4\textwidth]{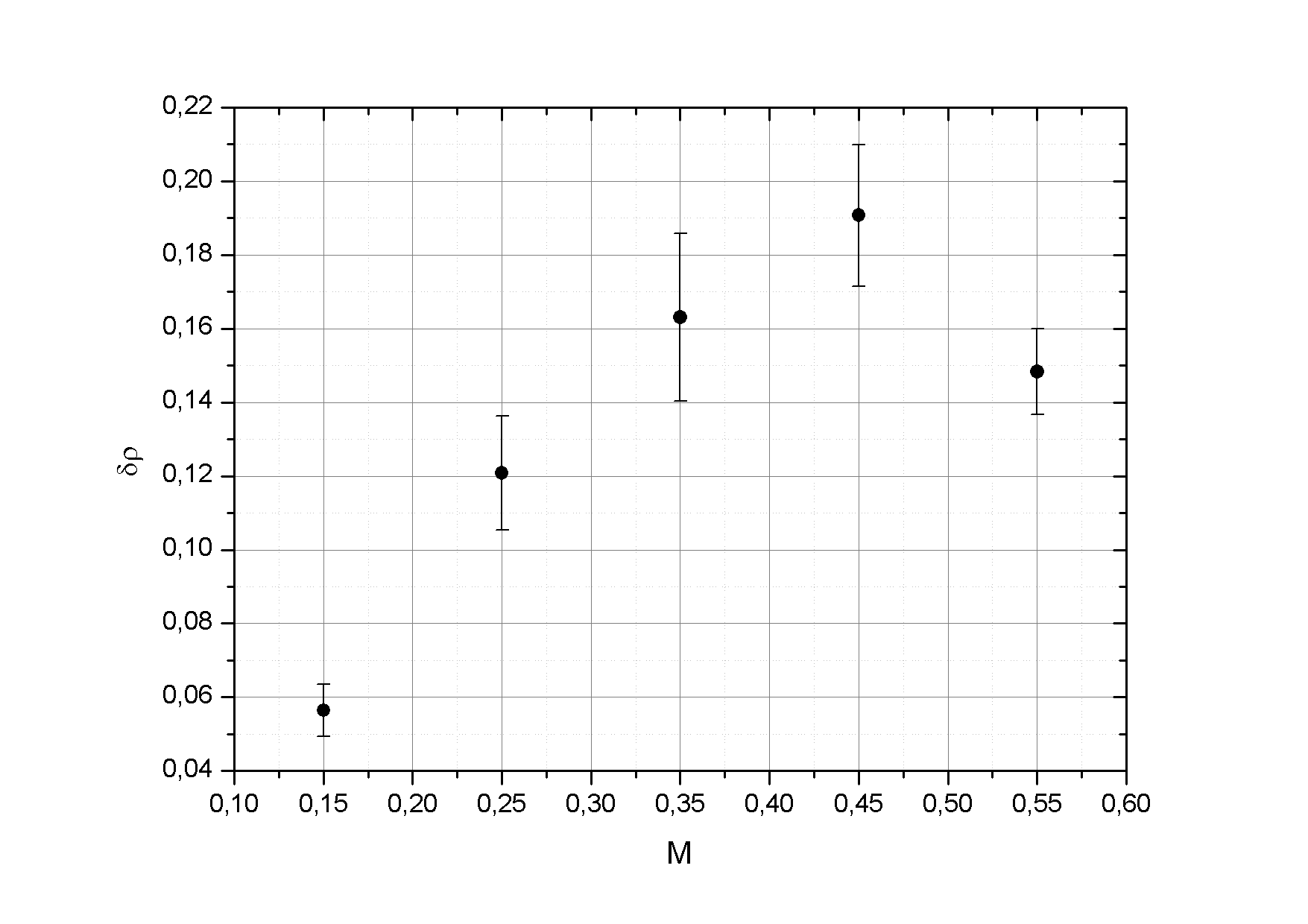}}
  \caption{\footnotesize {Density fluctuations vs Mach number, for compressible forcing.}}
  \label{fig:densidad-SI3}
\end{figure}

In figures \ref{fig:densidad-SI1}, \ref{fig:densidad-SI2} and \ref{fig:densidad-SI3} the density fluctuations are shown as a function of the Mach number for the simulations with incompressible, mixed and compressible forcing, respectively.

The value of the adjustment $\beta = 2.11 \pm 0.09$ for the simulations with incompressible forcing is compatible with the predictions of the NI theory, where it is predicted that density fluctuations must scale with the Mach number squared.
Through this analysis we can determine that for nearly incompressible systems with random initial conditions and incompressible forcing, we remain within the assumptions given by the nearly incompressible theory (NI).\\
The value of the adjustment $\beta = 1.00 \pm 0.04$ for the simulations with mixed forcing is compatible with the predictions of the MEC theory, where density fluctuations must scale with the Mach number. \\
Finally, the density fluctuations for compressible forcing seem to have a linear dependence with the Mach number for $M<0.5 $, then this dependence breaks down. These results, for Mach less than $0.5$, are consistent with previous works \cite{Ghosh} where it was found, by means of simulations in two-dimensions, that the density fluctuations scale linearly with the Mach number.

\subsection{Velocity fluctuations}

\begin{figure}[H]
\centerline{
  \includegraphics[width=0.4\textwidth]{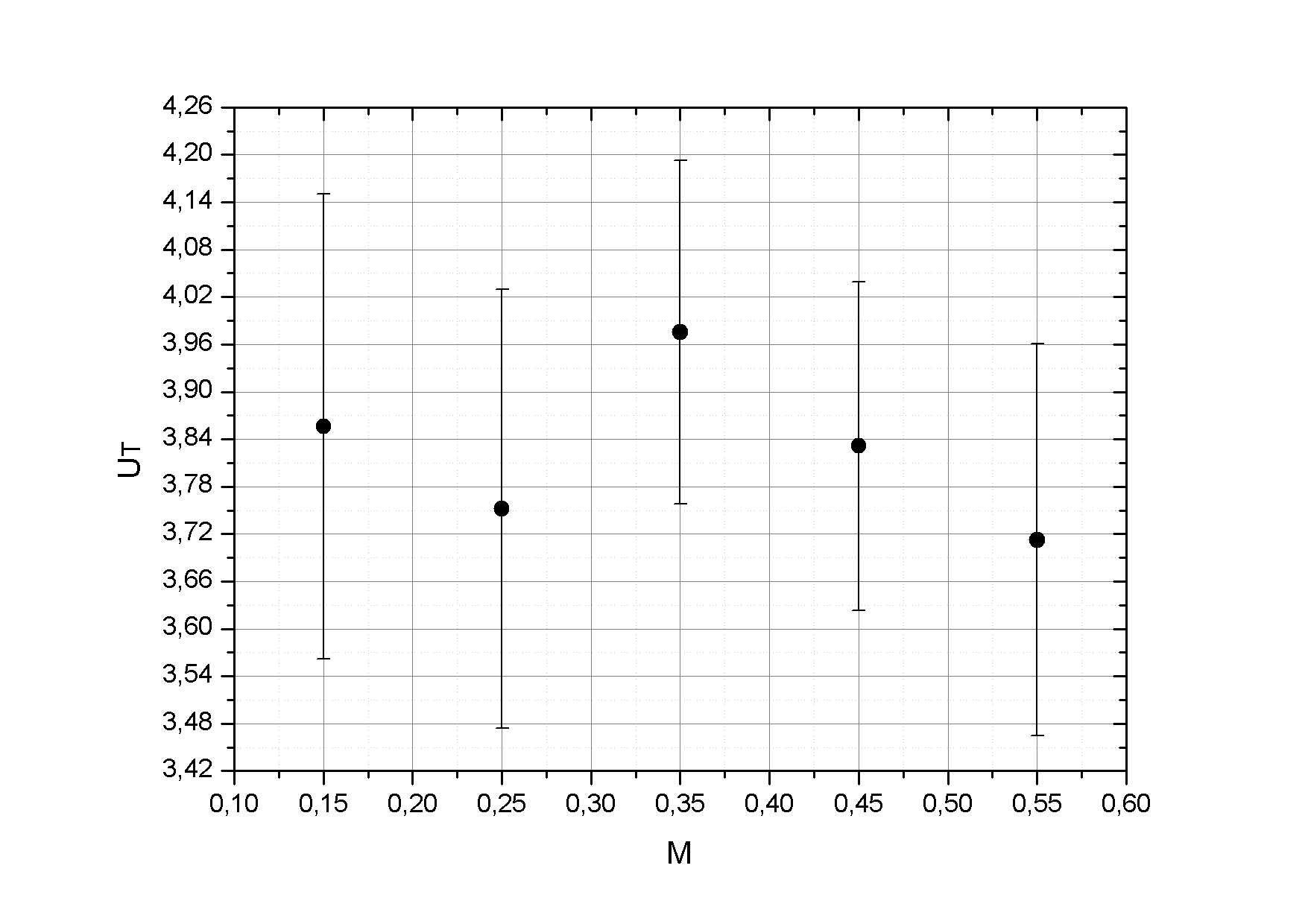}}
  \caption{\footnotesize {$U_{T}$ vs Mach number for incompressible forcing.}}
  \label{fig:UT-SI1}
\end{figure}

\begin{figure}[H]
\centerline{
  \includegraphics[width=0.4\textwidth]{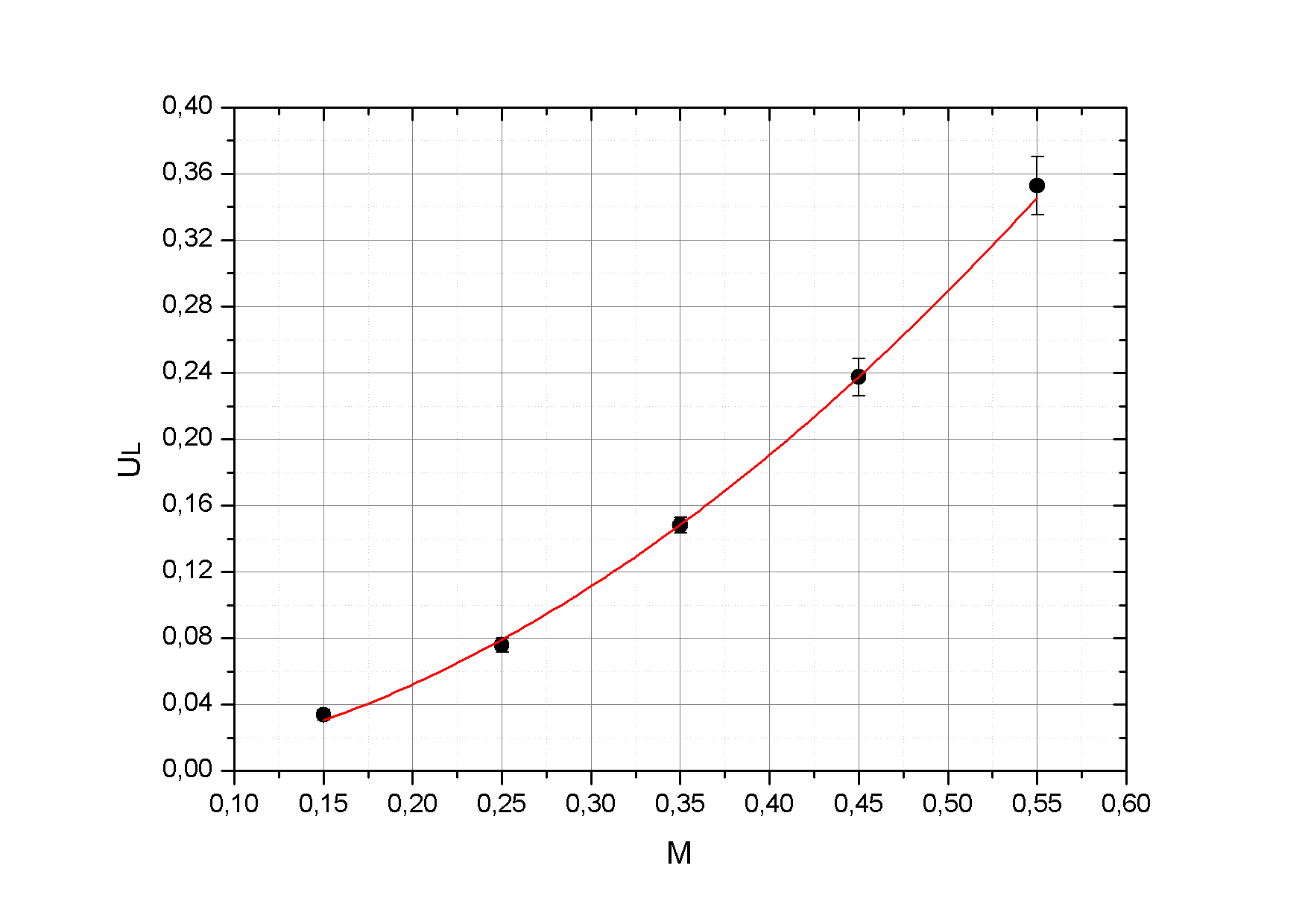}}
  \caption{\footnotesize {$U_{L}$ vs Mach number. The continuous line corresponds to the adjustment made for the simulations with incompressible forcing.}}
  \label{fig:UL-SI1}
\end{figure}

Similar to the density fluctuations, the NI, MEC, and CW theories have predictions for the ratio of longitudinal and perpendicular velocities.\\
The values of $U_{T}$, and $U_{L}$ vs Mach number for the incompressible forcing can be seen in Fig: \ref{fig:UT-SI1} and \ref{fig:UL-SI1}.

As we can see $U_{T}$ is always $\mathcal{O}(1)$ and is at least one order of magnitude
above $U_L$. Using a least square adjustment we determine that $U_{L}$ scales as $M^{1.87 \pm 0.06}$. This is consistent with the scaling of $\frac{\boldsymbol{\nabla} \cdot \textbf{u}}{\boldsymbol{\nabla} \times \textbf{u}}\sim \frac{U_{L}}{U_{T}}\sim M^{2}$ predicted by the NI theory.

\begin{figure}[H]
\centerline{
  \includegraphics[width=0.4\textwidth]{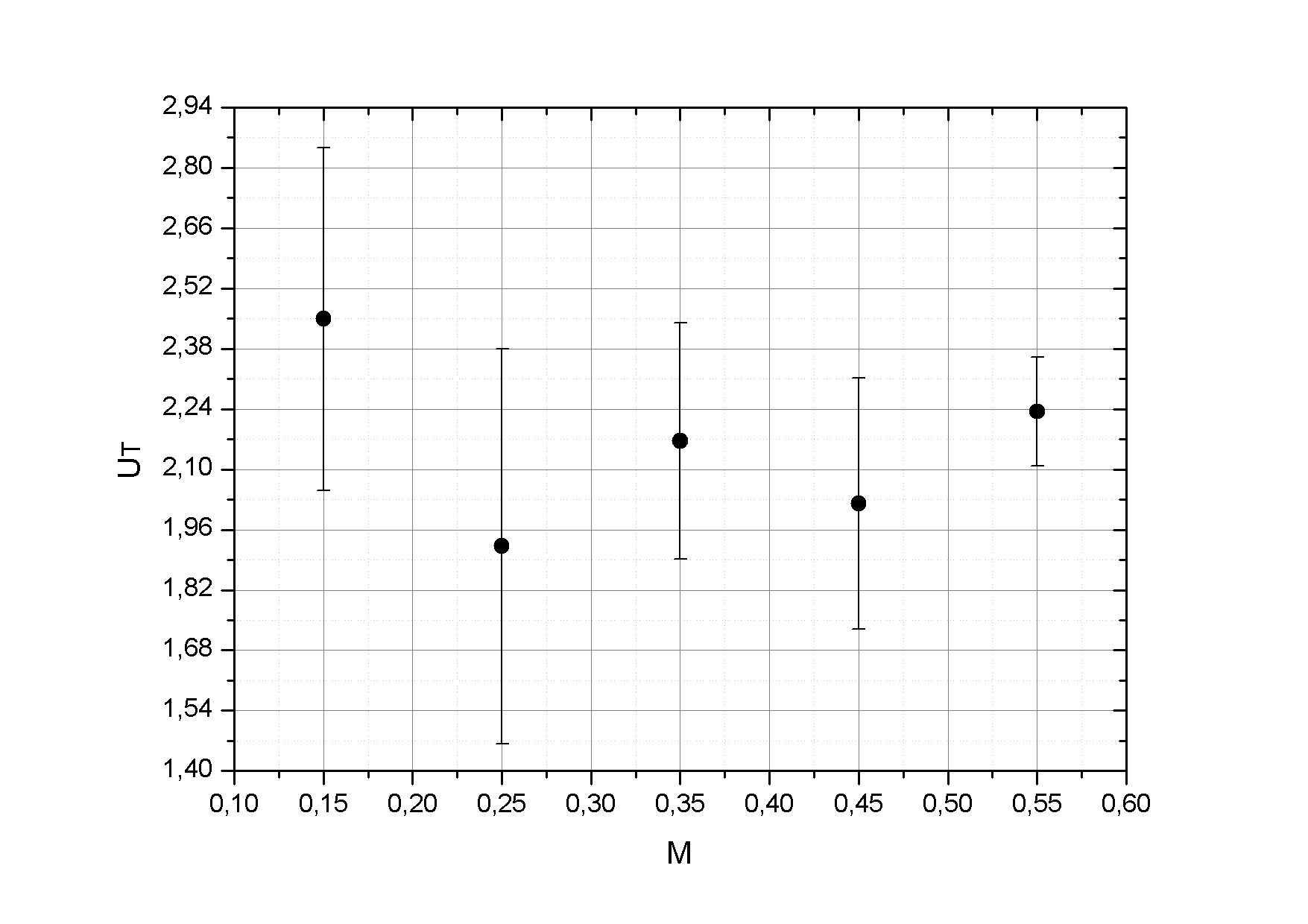}}
  \caption{\footnotesize {$U_{T}$ vs Mach number for mixed forcing.}}
  \label{fig:UT-SI2}
\end{figure}

\begin{figure}[H]
\centerline{
  \includegraphics[width=0.4\textwidth]{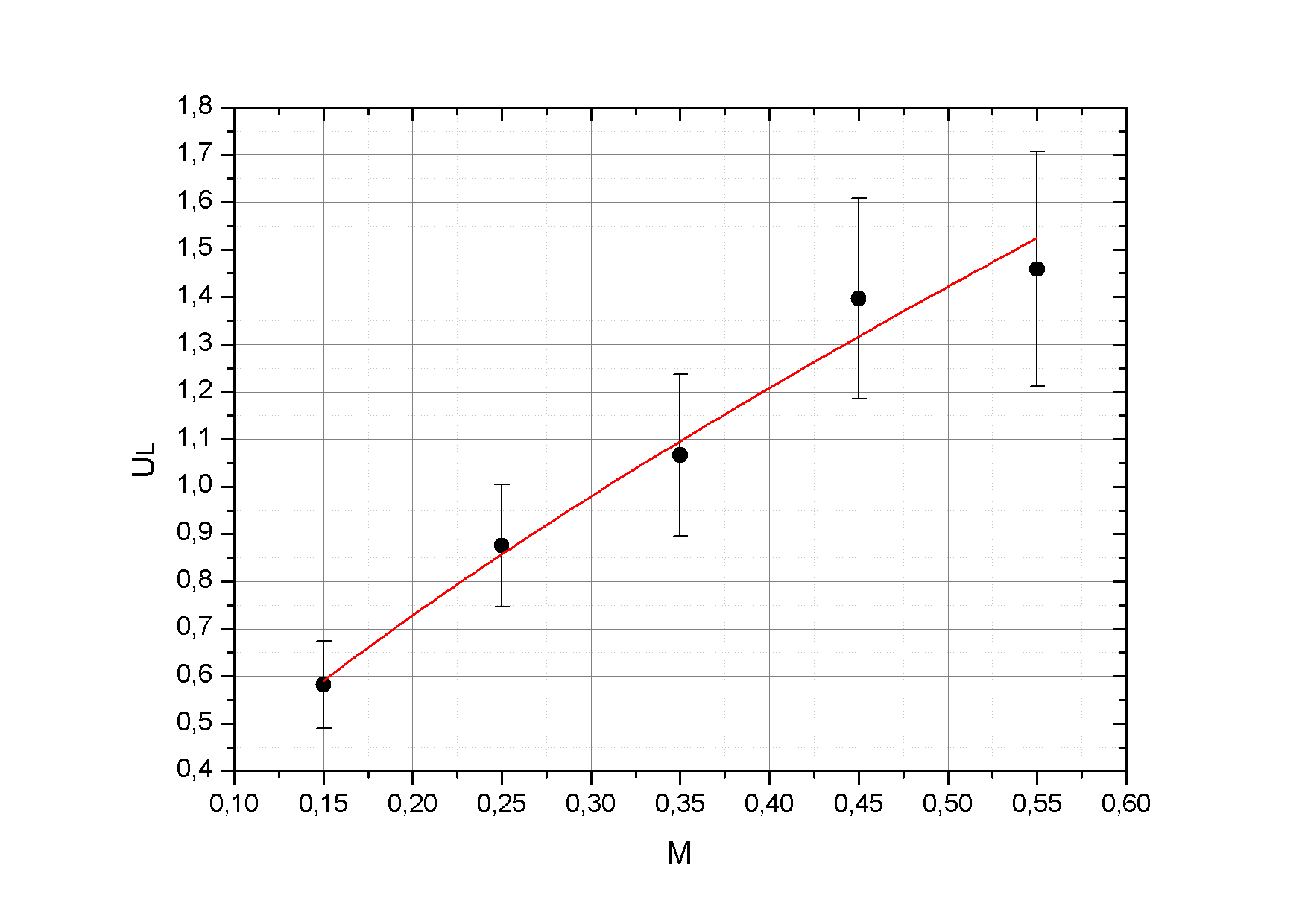}}
  \caption{\footnotesize {$U_{L}$ vs Mach number. The continuous line corresponds to the adjustment made for the simulations with mixed forcing.}}
  \label{fig:UL-SI2}
\end{figure}

Next we present the velocity fluctuations for mixed forcing in Fig: \ref{fig:UT-SI2} and \ref{fig:UL-SI2}.

In this case, $U_{T}$ is always $\mathcal{O}(1)$ and $U_{L}$ scales as $M^{0.73 \pm 0.05}$. For these simulations, $U_{L}$ and $U_{T}$ are almost always in the same order. 

\begin{figure}[H]
\centerline{
  \includegraphics[width=0.4\textwidth]{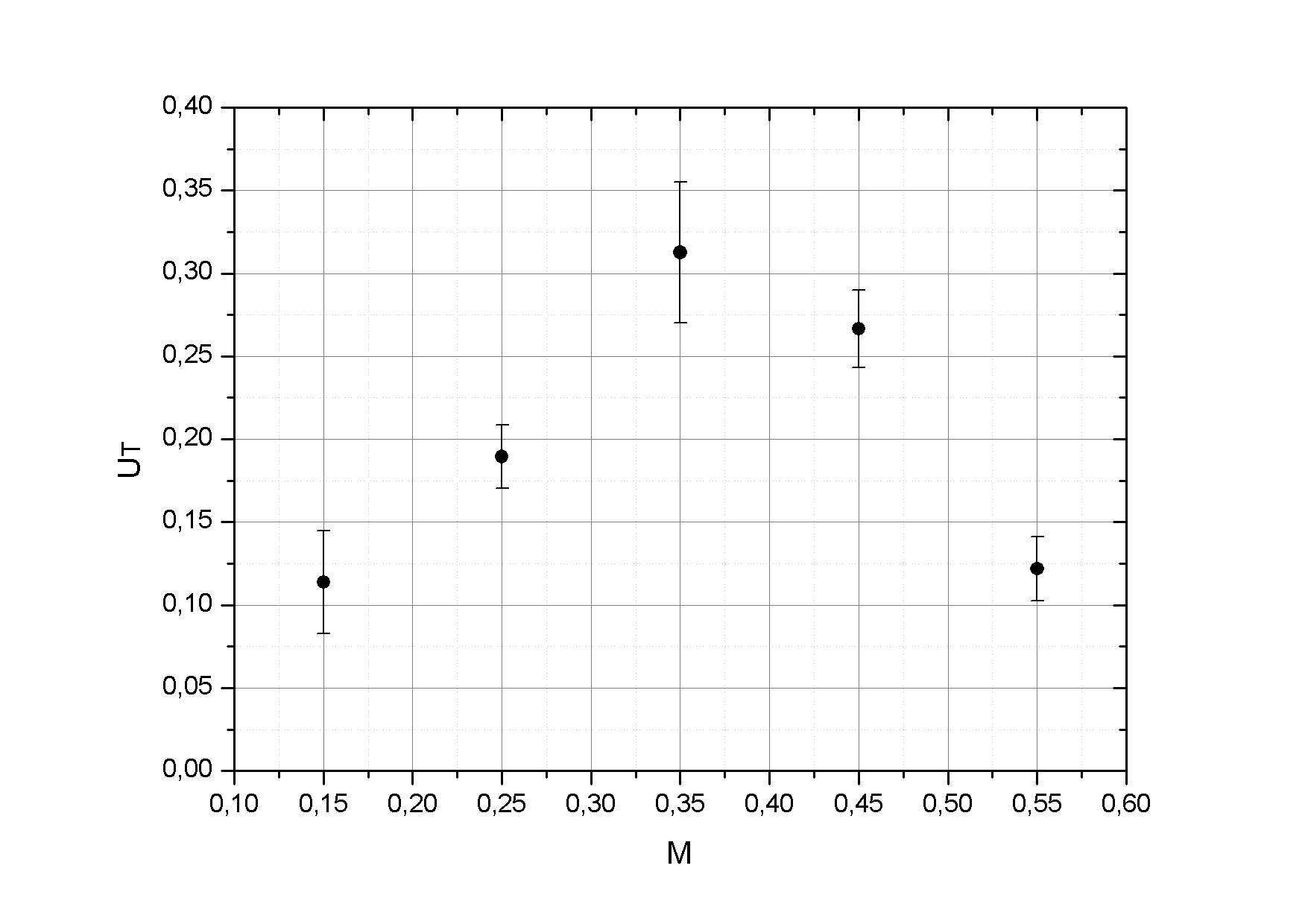}}
  \caption{\footnotesize {$U_{T}$ vs Mach number for compressive forcing.}}
  \label{fig:UT-SI3}
\end{figure}

\begin{figure}[H]
\centerline{
  \includegraphics[width=0.4\textwidth]{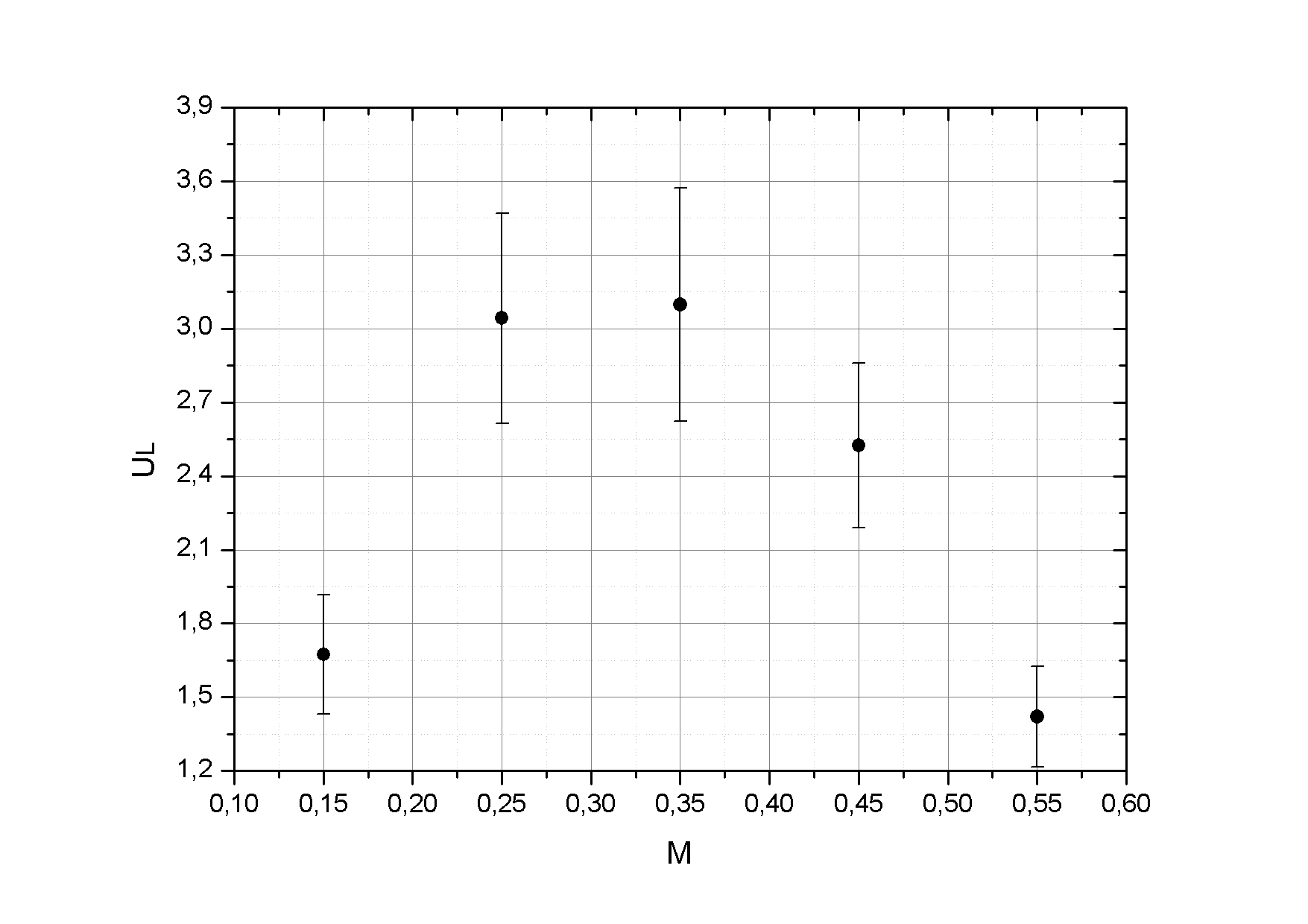}}
  \caption{\footnotesize {$U_{L}$ vs Mach number for compressive forcing.}}
  \label{fig:UL-SI3}
\end{figure}

Finally, we present the velocity fluctuations for compressive forcing in Fig: \ref{fig:UT-SI3} and \ref{fig:UL-SI3}.

For these simulations we find that both $U_{L}$ and $U_{T}$ does not present dependence on the Mach number and $U_{L}$ always dominates the dynamics of the system, being always at least an order of magnitude 
above $U_{T}$.

\subsection{Structures}

\begin{figure}[H]
\centerline{
  \includegraphics[width=0.4\textwidth]{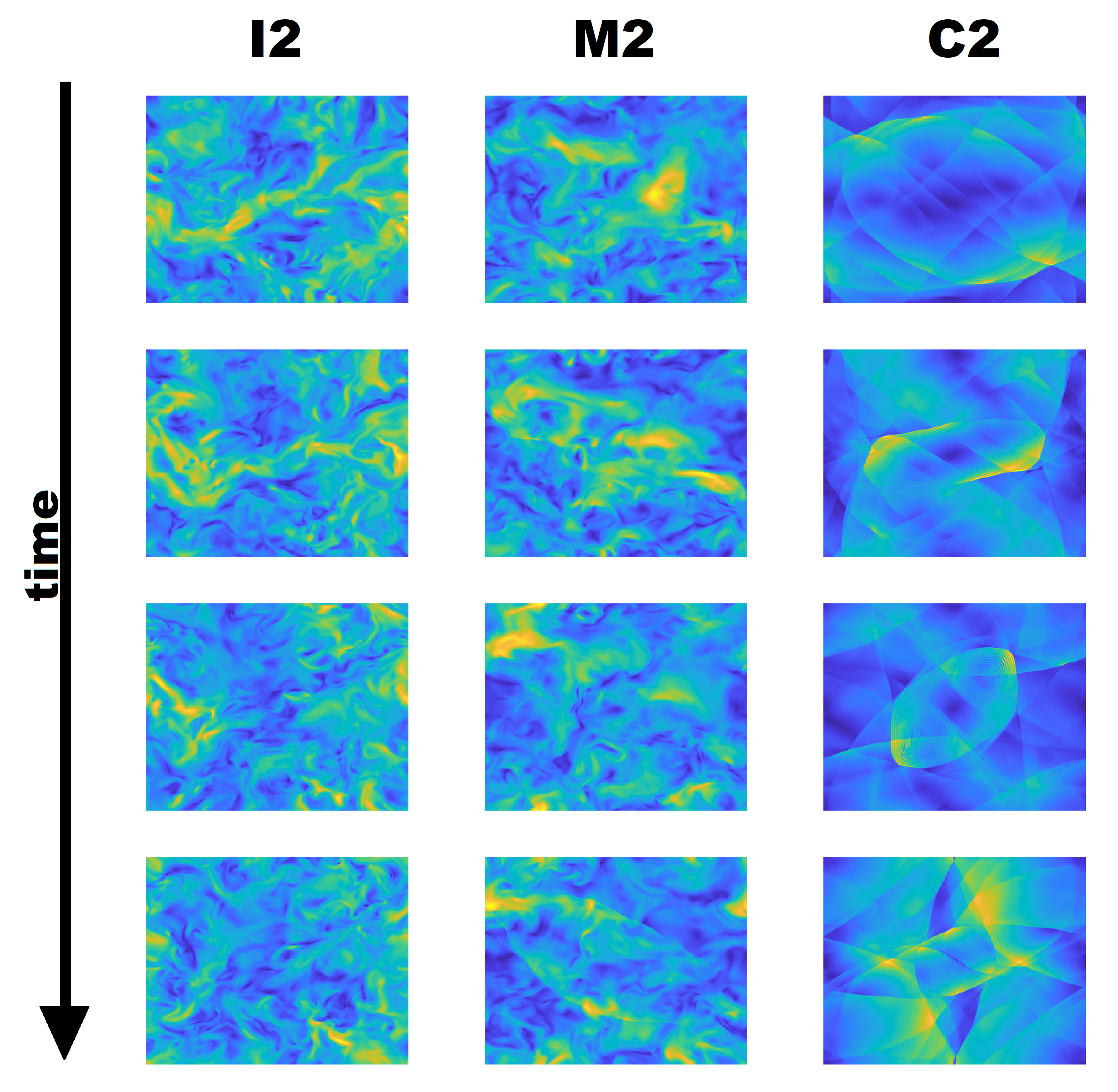}}
  \caption{\footnotesize {Cross section of the velocity field.}}
  \label{fig:est_v}
\end{figure}

\begin{figure}[H]
\centerline{
  \includegraphics[width=0.4\textwidth]{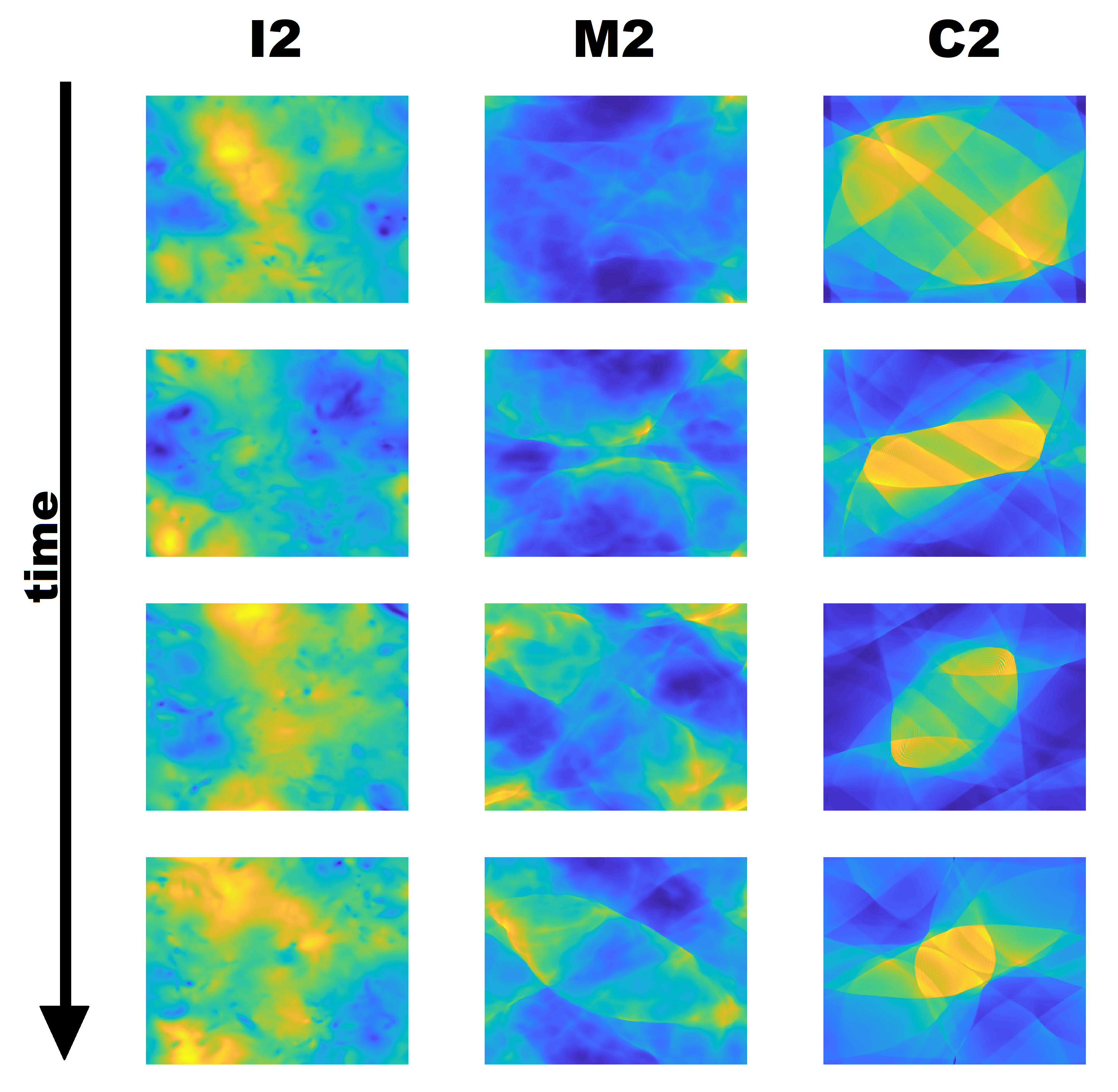}}
  \caption{\footnotesize {Cross section of the density field.}}
  \label{fig:est_rho}
\end{figure}

We show a cross section of the velocity field (See Fig: \ref{fig:est_v}) and density field (See Fig: \ref{fig:est_rho}) for a value of Mach number of $0.25$ and for the different types of forcing.

As we can see, the simulations with compressible forcing generate a number of significant large-scale structures, producing highly localized areas of high density. The velocity field follows a structure similar to that of the density field. We attribute this to the generation of shock waves by increasing the compressibility of the forcing in the fluid.

On the other hand, for mixed forcing type, some structures can be detected, although these are much less significant than those observed for compressible forcing. For incompressible forcing, the velocity and density field are homogeneous.

\section{\label{conclusions}Conclusions}

The studies between the limits of the near incompressible turbulence and the purely incompressible turbulence are an important step in the advance of the theory towards a complete treatment of the compressible flows.

Using models based on the recent advances in the mathematical structure of low-Mach fluids, we have described three perspectives on the nature of the near incompressible polytropic hydrodynamics.

We observe the generation of sound waves in the fluid, consistent with the known dispersion relation. We also observe different behaviors of the fluid at different scales, being the turbulence dominant in the small length scales (large $k$) and coexisting waves and turbulence at other scales, depending on the type of forcing applied. The concentration of energy around the dispersion relation of acoustic waves increases with the compressibility of the forcing.

For the energy spectra $ E(k)$, we find that they are compatible with the Kolmogorov prediction $k^{-5/3}$ for isotropic and homogeneous turbulence when using an incompressible forcing. For mixed and compressible forcing the spectra show a more pronounced energy decay compared to Kolmogorov law. We attribute this to extra dissipation channels. This seems plausible considering that in addition to the energy cascade from the large eddies to the small ones, there is also transfer from eddies to acoustic modes that are then quickly damped. The spectra present a similar profile for all values of the Mach number.

In addition, the axisymmetric spectra of kinetic energy were studied. These spectra show an energy concentration near the points with $k_{x}=k_{y}$, which becomes more evident as the compressibility of the forcing increases. These effects have not been observed in simulations or previous experiments under conditions of near incompressibility. The spectral anisotropy by increasing the compressibility of the forcing requires a more detailed analysis that is considered for study in future work. This effect is then reflected in the formation of large-scale structures, producing highly localized areas of high density. We attribute this to the generation of shock waves by increasing the compressibility in the fluid.

The temporal correlation function for each of the runs was studied. It was observed that the correlation times have, for an incompressible force, a change in behavior for high wave numbers ($k$) where the associated sweeping times predominate, while for low $k$ the time associated with acoustic waves 
dominates. This change in behavior strongly depends on the Mach number used. We assume that this behavior could become an effect of turbulent structure formation. On the other hand, for mixed and compressible forcing, the acoustic correlation times dominate the system in all the scales.

Finally, to corroborate the predictions described by NI and MEC theories and the antecedents observed in previous CW turbulence studies, density fluctuations and dependence on the Mach number was calculated, as well as the longitudinal and perpendicular velocity fluctuations. We observe consistency with the predictions of the NI, MEC and CW theories according to the type of forcing used, where systems with incompressible forcing seem to obey the scaling of fluctuations of density and velocities of the NI theory, while systems with mixed forcing obey the scaling of the MEC theory. The systems with compressible forcing seem to obey the turbulence CW previous studies for Mach values less than $0.5$. The velocity fluctuations seem not to be dependent on the Mach number.

The results of this work provide a certain degree of predictive power, at least within the framework of the specific model we have used. For example, by measuring fluctuations in density, the system can be unequivocally classified into one of the three groups we have described, and then making predictions about the rest of unmeasured magnitudes.

\section*{Acknowledgements}
The authors acknowledge support from PIP Grant No.~11220150100324CO and UBACYT Grant No. 20020170100508BA.

\end{document}